\documentclass[12pt]{article}
\usepackage{amsmath,amssymb,pstricks,pst-plot, pst-node, eucal, color}
\usepackage[dvips]{graphics}
\newtheorem{thm}{Theorem}[section]
\newtheorem{lemma}[thm]{Lemma}

\newcommand{\bh}{\bar h}
\newcommand{\R}{{\mathbb R}}
\newcommand{\E}{{\mathbb E}}
\newcommand{\C}{{\mathbb C}}
\newcommand{\Z}{{\mathbb Z}}
\newcommand{\D}{{\mathbb D}}
\newcommand{\T}{{\mathbb T}}
\newcommand{\Hy}{{\mathbb H}}
\newcommand{\ve}{\varepsilon}
\newcommand{\G}{{\cal G}}
\newcommand{\Gd}{{\cal G}_D}
\newcommand{\Gt}{{\cal G}_T}

\newcommand{\eps}{\epsilon}
\renewcommand{\H}{{\cal H}}
\newcommand{\Ht}{{\cal H}_T}
\renewcommand{\Re}{{\rm{Re}}}
\renewcommand{\Im}{{\rm{Im}}}

\renewcommand{\b}{\mathsf{b}}
\newcommand{\w}{\mathsf{w}}
\newcommand{\x}{\hat x}
\newcommand{\y}{\hat y}
\newcommand{\old}[1]{}
\newcommand{\GFF}{X}
\newcommand{\id}{\chi}

\newenvironment{proof}{\noindent{\bf Proof:} \hspace*{1em}}{
    \hspace*{\fill} $\square$\medskip }

\begin{document}
\title{Height fluctuations in the honeycomb dimer model}
\author{Richard Kenyon
\thanks{Department of Mathematics, University of British Columbia, Vancouver, B.C. Canada.}}
\date{}
\maketitle
\abstract{We study a model of random surfaces arising
in the dimer model on the honeycomb lattice. For a fixed ``wire frame''
boundary condition,
as the lattice spacing $\epsilon\to0$, Cohn, Kenyon and Propp \cite{CKP} 
showed the
almost sure convergence of a random surface to a 
non-random limit shape $\Sigma_0$. In \cite{KO},
Okounkov and the author showed how to parametrize the
limit shapes in terms of analytic functions, in particular
constructing a natural conformal structure on them.
We show here that when $\Sigma_0$ has no facets,
for a family of boundary conditions approximating the wire frame, 
the large-scale surface fluctuations
(height fluctuations) about $\Sigma_0$
converge as $\epsilon\to0$ to a Gaussian free field for the 
above conformal structure.
We also show that the local statistics
of the fluctuations near a given point $x$
are, as conjectured in \cite{CKP}, given by the unique ergodic Gibbs measure
(on plane configurations) whose slope is 
the slope of the tangent plane of $\Sigma_0$
at $x$.}

\tableofcontents

\section{Introduction}

\subsection{Dimers and surfaces}

A {\bf dimer covering}, or {\it perfect matching}, of a finite graph is a set of edges 
covering all the vertices exactly once. 
The {\bf dimer model} is the study of random dimer coverings of a graph.
Here we shall for the most part deal with the {\bf uniform} measure on dimer coverings.

In this paper we study
the dimer model on the {\bf honeycomb lattice}
(the periodic planar graph whose faces are regular hexagons), or rather,
on large pieces of it.
This model, and more generally dimer models 
on other periodic bipartite planar graphs,
are statistical mechanical models for discrete random interfaces.
Part of their interest lies in the
conformal invariance properties of their
scaling limits \cite{K.confinv,K.GFF}.

\begin{figure}
\begin{center}\scalebox{.5}{\includegraphics{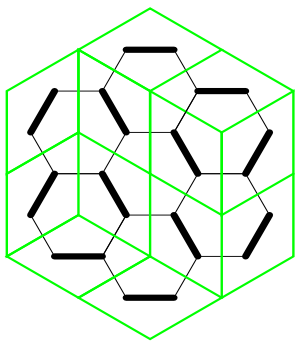}}
\end{center}
\caption{\label{dimerloz}Honeycomb dimers (solid)
and the corresponding ``lozenge''
tiling (green).}
\end{figure}

Dimer coverings of the honeycomb graph are dual
to tilings with $60^\circ$ rhombi,
also known as {\bf lozenges}, see Figure \ref{dimerloz}. 
Lozenge tilings can in turn be viewed as orthogonal projections onto the plane
$P_{111}=\{x+y+z=0\}$ of {\bf stepped surfaces} which
are polygonal
surfaces in $\R^3$ whose faces are squares in the $2$-skeleton of $\Z^3$
(the stepped surfaces are {\it monotone} in the sense 
that the projection is injective), see Figures \ref{dimerloz},\ref{bpp}.
Each stepped surface is the graph of a function, 
the {\bf normalized height function}, 
on the underlying tiling, which is linear on each tile. This function is 
defined simply as $\sqrt{3}$ times the distance from the surface to the plane $P_{111}$.
(The scaling factor $\sqrt{3}$ is just to make the function integer-valued on $\Z^3$.)

\subsection{Results}

We are interested in studying the {\bf scaling limit} of 
the honeycomb dimer model, that is, the limiting behavior of
a uniform random dimer covering of a fixed plane region $U$ when
the lattice spacing $\eps$ goes to zero. Equivalently, we take stepped
surfaces in $\eps\Z^3$ and let $\eps\to0$. As boundary conditions
we are interested in stepped surfaces spanning a ``wire frame'' which is a simple
closed polygonal path $\gamma_\eps$
in $\eps\Z^3$. We take $\gamma_\eps$ converging as $\eps\to0$ to a
smooth path $\gamma$ which projects to $\partial U$.

\subsubsection{Limit shape}

Let $U$ be a domain in $P_{111}$, and 
$\gamma$ be a smooth closed curve in $\R^3$, projecting orthogonally to $\partial U$.
For each $\eps>0$ let $\gamma_\eps$ be a nearest-neighbor path in
$\eps\Z^3$ approximating $\gamma$ 
(in the Hausdorff metric)
and which can be spanned by a
monotone stepped surface $\Sigma_\eps$, monotone in the sense
that it projects injectively to $P_{111}$, or in other words it 
is the graph of a function on $P_{111}$. See for example Figure \ref{bpp}
(although there the boundary is only piecewise smooth). 

The existence of such an approximating sequence imposes constraints on $\gamma$,
see \cite{CKP, Fournier}, as follows:
the curve $\gamma$ can be
spanned by a surface $\Sigma$, which is the graph of a continuous function
on $U$, and such that the normal $\nu$ to the surface points
into the positive orthant $\R_{\geq0}^3$. 
Conversely, any such curve $\gamma$ can be approximated by 
$\gamma_\eps$ as above, see \cite{CKP}. The condition of positivity of the normal to 
$\Sigma$ can be stated in terms of the gradient of the function $h$ whose
graph is $\Sigma$: this gradient must lie in a certain triangle. The formulation
in terms of the normal is more symmetric, however.

For a given $\gamma_\eps$ there are, typically, many spanning surfaces
$\Sigma_\eps$  and we
study the limiting properties of the uniform measure on the set of
$\Sigma_\eps$ as $\eps\to0$.

For a surface $\Sigma_\eps$ spanning $\gamma_\eps$,
let $h_\eps:P_{111}\to\R$ be the normalized height function,
defined on the region enclosed by $U_\eps:=\pi_{111}(\gamma_\eps)$, whose
graph is $\Sigma_\eps$.

Under the above hypotheses Cohn, Kenyon and Propp proved the existence
of a limit shape:
\begin{thm}[Cohn, Kenyon, Propp \cite{CKP}]\label{ckpthm} The distribution of 
$h_{\eps}$ converges as $\eps\to0$ a.s. to a {\it nonrandom} function $\bh:U\to\R$. 
The function $\bh$
is the unique function $h$ which minimizes the ``surface tension''
functional 
$$\min_h\int_U \sigma(\nabla h) \,dx\,dy,$$
where, in terms of the normal vector $(p_a,p_b,p_c)\in\R^3$ to the graph of $h$
scaled so that $p_a+p_b+p_c=1$, we have
$\sigma(p_a,p_b,p_c)=-\frac1{\pi}(L(\pi p_a)+L(\pi p_b)+L(\pi p_c))$ and
$L(x)=-\int_0^x\log 2\sin t\,dt$ is the Lobachevsky function.
\end{thm}

Here the minimum is over Lipschitz functions whose graph has normal 
with nonnegative coordinates. Equivalently, these are functions whose gradient
lies in a certain triangle. In the above formula the surface tension $\sigma$ 
is the negative of the exponential growth rate of the number
of discrete surfaces of average slope $\nabla h$.

The function $\bh$ is called the {\bf asymptotic height function}.
Its graph is a surface $\Sigma_0$ spanning $\gamma$.

\subsubsection{Local statistics}

Suppose that the gradient
of $\bh$ is not maximal at any point in $\bar U$, i.e.
the normal to $\Sigma_0$ has nonzero coordinates at every point of $\bar U$.
In this paper we show that, if the precise local behavior of
the approximating curves $\gamma_\eps$ 
is chosen in a particular way, then both the local statistics and the global
height fluctuations of $\Sigma_\eps$ can be determined. 

Here is the result on the local statistics.
\begin{thm}\label{main0}
Suppose that the gradient
of $\bh$ is not maximal at any point in $\bar U$, that is, the normal vector
to the surface has nonzero coordinates at every point. Under appropriate 
hypotheses on the local structure of the approximating curves $\gamma_\eps$,
the local statistics of $\Sigma_\eps$ near a given point
are given by the unique Gibbs measure of slope equal to the 
slope of $\bh$ at that point.
\end{thm}

For the precise statement see Theorem \ref{Kinvasymp}.
In particular the hypotheses on $\gamma_\eps$ are explained in
sections \ref{Tgraphsection} and \ref{boundarybehavior}.

Suppose that $(p_a,p_b,p_c)$
is a normal vector to the surface at a point. Recall that $p_a,p_b,p_c>0.$ 
If we rescale so that $p_a+p_b+p_c=1$, then one consequence of 
Theorem \ref{Kinvasymp} is that the quantities $p_a,p_b,p_c$ are the densities
of the three orientations of lozenges near the corresponding point on the surface.

\subsubsection{Fluctuations}\label{abcdef}

\begin{figure}
\begin{center}\scalebox{.7}{\includegraphics{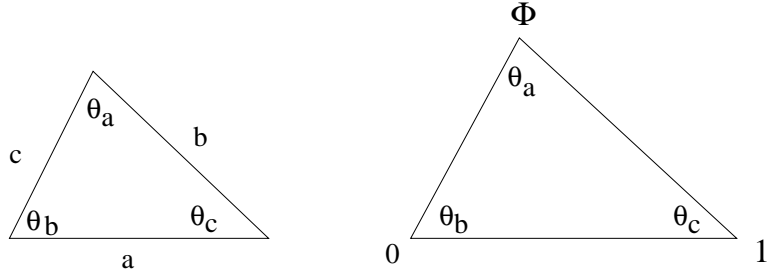}}\end{center}
\caption{\label{triangles}The triangle and a scaled copy with vertices
$0,1,\Phi$.}
\end{figure}

The fluctuations are the image of the Gaussian free field under a 
certain diffeomorphism from the unit disk $\D$ to $U$.
To describe the fluctuations, we first describe the relevant conformal structure on $U$.
It is a function of the normal to the graph of $\bh$, and is defined as follows.
Let $(p_a,p_b,p_c)$ be the normal to $\bh$, scaled so that $p_a+p_b+p_c=1$.
Let $\theta_a=\pi p_a,\theta_b=\pi p_b,\theta_c=\pi p_c$.
Let $a,b,c$ be the edges of a Euclidean triangle with angles
$\theta_a,\theta_b,\theta_c$.
Let $z=-e^{-i\theta_c}$ and $w=-e^{i\theta_b}$ so that $a+bz+cw=0$.
See Figure \ref{triangles}; here the triangle on the left has edges $a,bz,cw$ when
these edges are oriented counterclockwise.
We define $\Phi=-cw/a$.
All of these quantities are functions on $U$, although $a,b,c$ 
are only defined up to scale.
Let $\hat x,\hat y,\hat z$ be the unit vectors in $P_{111}$
in the directions of the projections of the 
standard basis vectors in $\R^3$.

\begin{thm}[\cite{KO}]\label{burgers}
The function $\Phi$ satisfies the complex Burgers equation
\begin{equation}\label{pde}
\Phi_{\x}+\Phi \Phi_{\y}=0,
\end{equation}
where $\Phi_{\x},\Phi_{\y}$ are directional derivatives of $\Phi$ in directions
$\x,\y$ respectively.
\end{thm}
 
The function $\Phi:U\to\C$ can be used to define a conformal structure
on $U$, as follows. A function $g:U\to\C$ is defined to be analytic in this
conformal structure on $U$ if it satisfies
$g_{\x}+\Phi g_{\y}=0$, where $g_{\x},g_{\y}$ are the directional derivatives
of $g$ in the directions $\hat x,\hat y$ respectively.
By the Alhfors-Bers theorem there is a diffeomorphism $f:U\to\D$
satisfying $f_{\x}+\Phi f_{\y}=0$; the conformal structure on $U$ is the 
pull-back of the standard conformal structure on $\D$ under $f$.
The conformal structure on $U$ can be described by a {\it Beltrami coefficient} $\xi$
(see below) which in the current case is $\xi=(\Phi-e^{i\pi/3})/(\Phi-e^{-i\pi/3})$.

In the special cases that $\Phi$ is constant
(which correspond to the cases where $\gamma$ is contained in a plane),
this means that the conformal structure is just a linear image of
the standard conformal structure. For example, note that in the standard conformal
structure on $P_{111}$, a function $g$ is analytic if $g_{\x}+e^{i\pi/3}g_{\y}=0$.
So the case $\Phi=e^{i\pi/3}$, which corresponds to the case $a=b=c$,
gives the standard conformal structure
(recall that the vectors $\x$ and $\y$ are $120^\circ$ apart).

\begin{thm}\label{main1}
Suppose that the gradient
of $\bh$ is not maximal at any point in $\bar U$. Under the same hypotheses on $\gamma_\eps$ as in Theorem \ref{main0}, the fluctuations of the {\it
unnormalized} height function, $\frac1{\eps}(h_\eps-\bh)$, have a
weak limit as $\eps\to0$ which is the Gaussian free field
in the complex structure defined by $\Phi$, that is, the pull-back under
the map $f:U\to\D$ above, of the Gaussian free field on the unit disk $\D$.
\end{thm}
 
For the definition of the Gaussian free field see below.

As mentioned above, 
we require that the normal to the graph of $\bh$
be {\it strictly} inside the positive orthant, so that we have a positive
lower bound on the values of $p_a,p_b,p_c$, whereas the results of \cite{CKP} and
\cite{KO}
do not require this restriction. Indeed, in many of the simplest cases 
the surface $\Sigma_0$ will have facets, which are regions on which $(p_a,p_b,p_c)=(1,0,0),(0,1,0)$ or $(0,0,1)$. Our results do not apply to these situations.

This work builds on work of \cite{K.GFF,KOS,KO}. 
Previously Theorem \ref{main1}
was proved for a the dimer model on $\Z^2$
in a special case which in our context corresponds to the wire frame $\gamma$
lying in the plane $P_{111}$, see
\cite{K.GFF}. In that case
the conformal structure is the standard conformal structure on
$U$.

In the present case we still require special boundary conditions,
which generalize the ``Temperleyan" boundary conditions of
\cite{K.confinv, K.GFF}. It remains an open question
whether the result holds for all boundary conditions. 
The fluctuations in the presence of facets are also unknown,
and the current techniques to not seem to immediately
extend to this more general setting.

\subsection{The Gaussian free field}

The Gaussian free field $\GFF$ on $\D$
\cite{Sheff.GFF} is a random object in the space of distributions on $\D$, 
defined on smooth test functions as follows.
For any smooth test function 
$\psi$ on $\D$, $\int_\D \psi(x)\GFF(x)|dx|^2$  is a real Gaussian
random variable of mean zero and variance given by 
$$\int_{\D}\int_{\D}\psi(x_1)\psi(x_2)G(x_1,x_2)|dx_1|^2|dx_2|^2,$$
where the kernel $G$ is the Dirichlet Green's function on $\D$:
$$G(x_1,x_2) = -\frac1{2\pi}\log\left|\frac{x_1-x_2}{1-\bar x_1x_2}\right|.$$
%The Gaussian free field
%is conformally invariant, when thought of as a (signed) measure:
%that is, if $\phi:V\to \D$ is a conformal diffeomorphism, then we can define 
%the GFF $\GFF_V$ on $V$ by
%$$\int_V\psi(x)\GFF_V(x)|dx|^2 = \int_\D \psi(\phi(x))\GFF(\phi(x))|\phi'(x)|^2|dx|^2.$$

A similar definition holds (for the standard conformal structure) on any 
bounded domain in $\C$,
only the expression for the Green's function is different.

An alternative description of the Gaussian free field is that it is
the unique Gaussian process which satisfies
$\E[\GFF(x_1)\GFF(x_2)]=G(x_1,x_2)$.
Higher moments of Gaussian processes can always be written in terms of 
the moments of order $2$; for the Gaussian free field we have
$$\E(\GFF(x_1)\dots\GFF(x_{n}))=0 \text{~~~if $n$ is odd,}$$
and
\begin{equation}\label{moms}
\E(\GFF(x_1)\dots\GFF(x_{2k}))=
\sum_{\text{pairings}}G(x_{\sigma(1)},x_{\sigma(2)})\dots G(x_{\sigma(2k-1)},x_{\sigma(2k)})
\end{equation}
where the sum is over all $(2k-1)!!$ pairings of the indices.
Any process whose moments satisfy (\ref{moms}) is the Gaussian free field
\cite{K.GFF}.

\subsection{Beltrami coefficient}

A conformal structure on $U$ can be defined as an equivalence class of
diffeomorphisms $\phi:U\to\D$, where 
mappings $\phi_1,\phi_2$ are equivalent if the composition
$\phi_1\circ\phi_2^{-1}$ is a conformal self-map of $\D$.
The {\bf Beltrami differential}
$\xi(z)\frac{d\bar z}{dz}$ of $\phi$ is defined by the formula
$$\xi(z)\frac{d\bar z}{dz}=\frac{\phi_{\bar z}}{\phi_z} \frac{d\bar z}{dz}.$$
The Beltrami differential is invariant under post-composition of $\phi$ with
a conformal map, so it is a function only of the conformal structure
(and in fact defines the conformal structure as well). 
It is not hard to show that $|\xi(z)|<1$;
note that $\xi(z)=0$ if and only if the map is conformal.
The Ahlfors-Bers uniformization theorem \cite{AB} says that any smooth function
(even any measurable function) $\xi(z)$ satisfying $|\xi(z)|<1$ defines a conformal
structure.

\subsection{Examples}

The simplest case is when the wire frame $\gamma$ is contained in a plane
$\{(x,y,z)\in{\R^3}~|~p_a x+p_b y+p_c z=const\}$.
In this case the limit surface $\Sigma_0$ is linear. 
The normal $(p_a,p_b,p_c)$ is constant,
and the 
conformal structure is a linear
image of the standard conformal structure. That is, the map $f:U\to\D$
is a linear map $L$ composed with a conformal map.

\begin{figure}
\begin{center}\scalebox{.8}{\includegraphics{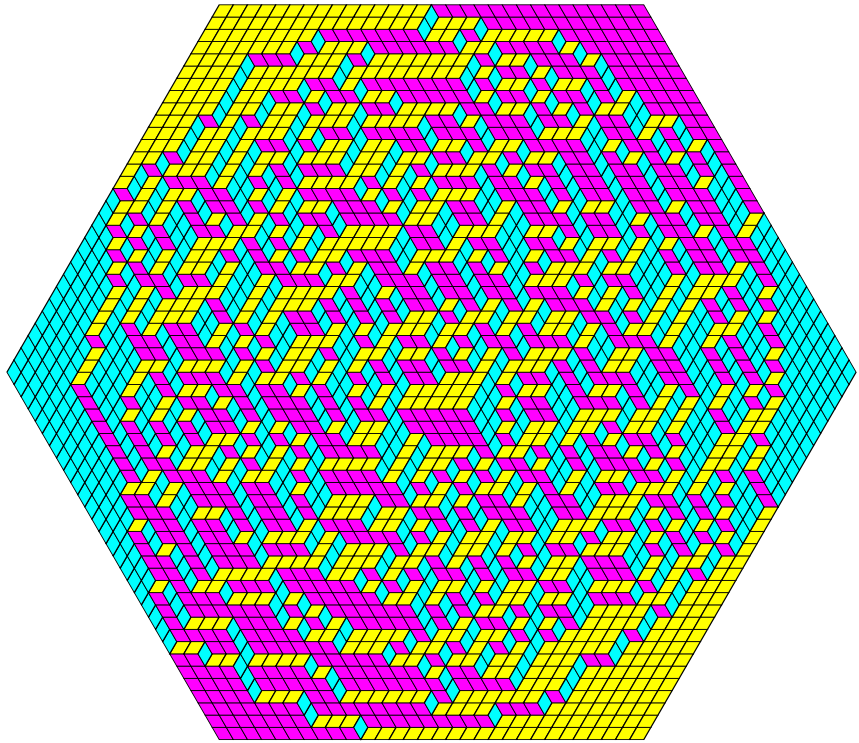}}\end{center}
\caption{\label{bpp}Boxed plane partition.}
\end{figure}

For a more interesting case, consider the {\bf boxed plane partition} (BPP)
shown in Figure \ref{bpp}, which is a random lozenge tiling of a regular hexagon. 
In \cite{CLP} it was shown that, for a random tiling
of the hexagon, the asymptotic height function $\bh$ 
is linear outside of the inscribed
circle and analytic inside (with an explicit but somewhat complicated formula).
Although our theorem does not apply to this case because of the
facets outside the inscribed circle,
if we choose boundary conditions inside the inscribed circle, 
and boundary values equal to the graph of the function there,
our results apply.
Suppose that the hexagon has sides of length $1$, so that
the inscribed circle has radius $\sqrt{3}/2$.
Let $U$ be a disk of radius $r<\sqrt{3}/2$ concentric with it.
Suppose that  the normalized height 
function on the boundary of $U$ is chosen to agree with the asymptotic 
height function of the
corresponding region in the BPP, so that the asymptotic height
function of $U$ equals the asymptotic height function of the BPP restricted to $U$.
Then the fluctuations on $U$ can be computed using Theorem \ref{main}. 
In fact in this setting, the conformal structure can be explicitly computed: 
take the standard conformal structure on
a hemisphere in $\R^3$, and project it orthogonally onto the plane containing
its equator. Identifying the equator with the inscribed circle in the BPP gives the 
relevant conformal structure in $U$.
Remarkably, in this example
the Beltrami coefficient is rotationally invariant, even though $\bh$ itself is not. 
See section \ref{BPPsection}.

We conjecture that the fluctuations for the BPP are given by the limit
$r\to\sqrt{3}/2$ of this construction (it is known \cite{CLP} 
that fluctuations
in the ``frozen'' regions outside the circle are exponentially 
small in $1/\eps$).

\subsection{Proof outline}

The fundamental tool in the study of the dimer model is the 
Kasteleyn matrix (defined below). Minors of the inverse Kasteleyn
matrix compute edge correlations in the model.
The main goal of the paper is to obtain an asymptotic expansion of the inverse
Kasteleyn matrix. This is complicated by the fact that it grows
exponentially in the distance between vertices 
(except in the special case when the boundary height function is
horizontal). However by pre- and post-composition with an 
appropriate diagonal matrix, we can remove the exponential
growth and relate $K^{-1}$ to the standard
Green's function with Dirichlet boundary conditions on a related graph $\Gt$.

Here is a sketch of the main ideas.
\begin{enumerate}
\item We construct a discrete version of the map $f$ of Theorem \ref{main1}.
For each $\eps$ we define a directed graph $\Gt$
embedded in the upper half plane $\Hy$, and a geometric map
$\phi$ from $U_\eps$ to $\Gt$, 
such that the Laplacian on $\Gt$ is related (via the construction of \cite{KPW,KS})
with the Kasteleyn matrix on $U_\eps$.
The existence of such a graph $\Gt$ follows from \cite{KS}.
This is done in section \ref{GT.CS} in the ``constant slope'' case and
section \ref{gen.CS} in the general case.

\item Standard techniques for discrete harmonic functions yield an 
asymptotic expansion 
for the Green's function on $\Gt$. This is done in section \ref{CS.green}.

\item The asymptotic expansion 
of the inverse Kasteleyn matrix
on $U_\eps$ is obtained from the derivative of the Green's function on $\Gt$, 
pulled back under the mapping $\phi$. See sections \ref{CS.Kinv} and \ref{gen.green}.

\item Asymptotic expansions 
of the moments of the height fluctuations are computed
via integrals of the asymptotic inverse Kasteleyn matrix. These moments
are the moments of the Gaussian free field on $\Hy$ pulled back under $\phi$.
\end{enumerate} 
\medskip

\noindent{\bf Acknowledgments.}
Many ideas in this paper were inspired by conversations
with Henry Cohn, Jim Propp, Jean-Ren\'e
Geoffroy, Scott Sheffield, 
B\'eatrice deTili\`ere, C\'edric Boutillier, and Andrei Okounkov.
We thanks the referees for useful comments.
This paper was partially completed while the author was visiting Princeton University.

\section{Definitions}

\subsection{Graphs}\label{graphs}

\begin{figure}[htbp]
\begin{center}\scalebox{.7}{\includegraphics{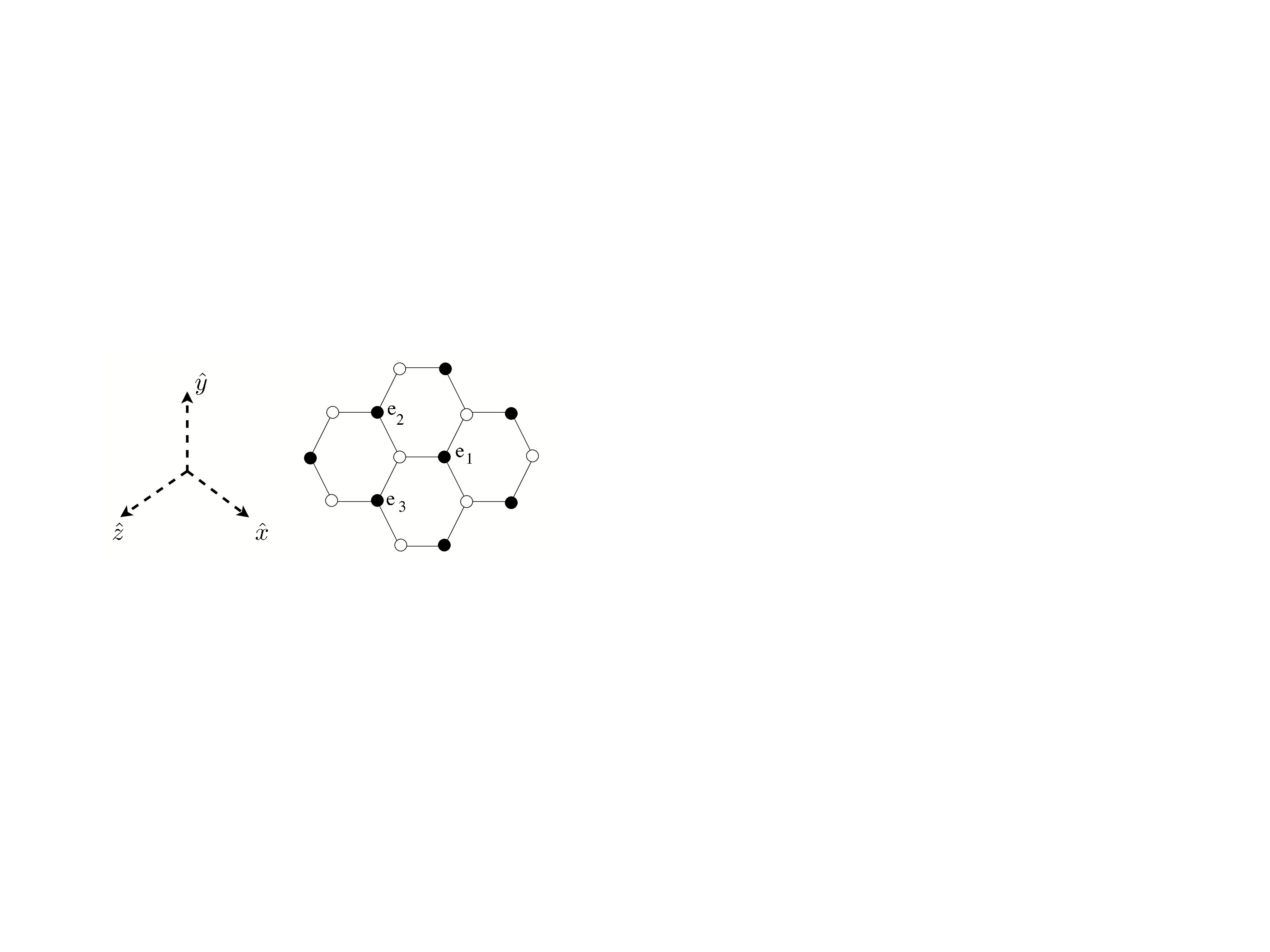}}
\end{center}
\caption{\label{honeycomb}The honeycomb graph.}
\end{figure}

Let $\pi_{111}$ be the orthogonal projection of $\R^3$ onto
$P_{111}$. Let $\hat x,\hat y,\hat z$ be the 
$\pi_{111}$-projections of the unit basis vectors. Define $e_1=\frac13(\hat x-\hat z),
e_2=\frac13(\hat y-\hat x),e_3=\frac13(\hat z-\hat y),$
so that $\hat x=e_1-e_2,\hat y=e_2-e_3,\hat z=e_3-e_1$.
Let $\H$
be the honeycomb lattice in $P_{111}$: vertices of $\H$ are $L\cup
(L+e_1)$, where $L$ is the lattice $L=\Z(e_1-e_2)+\Z(e_2-e_3)=
\Z\hat x+\Z\hat y,$ 
and edges connect nearest neighbors. 
Vertices in $L$ are colored white, those in $L+e_1$ are black.
See Figure \ref{honeycomb}.

\subsubsection{Dual graph}\label{dualgraph}

\begin{figure}[htbp]
\begin{center}\scalebox{.8}{\includegraphics{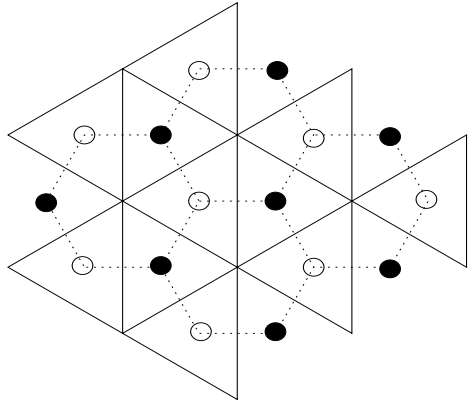}}
\end{center}
\caption{\label{G*}The ``dual" graph $\G^*$ (solid lines) of the graph $\G$ of 
Figure \ref{honeycomb}.}
\end{figure}

Let $U$ be a Jordan domain in $P_{111}$ with smooth boundary. 
In $\eps\H$, take a simple closed polygonal path with approximates
$\partial U$ in a reasonable way, for example the polygonal curve
is locally monotone in the same direction as the curve $\partial U$.
Let $\G$ be the subgraph of $\eps\H$
bounded by this polygonal path. 
We define a special kind of dual graph $\G^*$
as follows. Let $\eps\H^*$ be the usual planar dual of $\eps\H$. For each
white vertex of $\G$ take the corresponding triangular face of $\eps\H^*$; the
union of the edges forming these triangles, along with the
corresponding vertices, forms $\G^*$. In other words, $\G^*$ has a
face for each white vertex of $\G$, as well as for black vertices
which have all three neighbors in $\G$. See Figure \ref{G*}.

Throughout the paper the graph $\G$ and its
related graphs will be scaled by a factor $\eps$ over the corresponding
graphs $\H$, and so the $\G$ graphs have edge lengths of order $\eps$, and the graph
$\H$ and its related graphs have edge lengths of order $1$.

\subsubsection{Forms}

For an edge in $\G$ joining vertices $\b$ and $\w$, we denote by
$(\b\w)^*$ the dual edge in $\G^*$, which we orient at $+90^\circ$ from the edge
$\b\w$ (when this edge is oriented from $\b$ to $\w$).

A {\bf $1$-form} $\omega$ on a graph is a function on directed edges which is
antisymmetric with respect to reversing the orientation: 
$\omega(v_1v_2)=-\omega(v_2v_1)$. A $1$-form is also called a flow.
If the graph is planar one can similarly define a $1$-form on the dual graph. 
If $\omega$ is a $1$-form, $\omega^*$ is the dual $1$-form, defined by $\omega^*((v_1v_2)^*):=\omega(v_1v_2)$.

On a planar graph with a $1$-form $\omega$, $d\omega$ is a function on
oriented faces defined by $d\omega(f) = \sum_e \omega(e)$ where the 
sum is over the edges on a path around the face going counterclockwise.
This is also known as the curl of the flow $\omega$. 
The form $d\omega^*$ is a function on vertices 
(faces of the dual graph), defined by
$d\omega^*(v)=\sum_{v'\sim v}\omega(v,v')$. In other words it is the
divergence of the flow $\omega$.

A $1$-form $\omega$ is {\bf closed} if $d\omega=0$, that is, 
the sum of $\omega$ along any cycle is zero (in the language of flows,
the flow has zero curl). If is {\bf exact} if
$\omega=df$ for some function $f$ on the vertices, that is
$\omega(v_1v_2)=f(v_1)-f(v_2).$ 
A $1$-form is {\bf co-closed} if its dual form is closed. The 
corresponding flow is divergence-free. 

If $d\omega^*=0$, the integral of $\omega^*$ between two faces of $\G$
(i.e. on a path in the dual graph) is the {\bf flux}, or total flow, between those faces.

\subsection{Heights and asymptotics}\label{abcdefs}

The {\bf unnormalized} height function, or just height function,
of a tiling is the integer-valued function on the vertices of the
lozenges (faces of $\G$) 
which is the sum of the coordinates of the corresponding
point in $\Z^3$. It changes by $\pm1$ along each edge of a tile. 
When we scale the lattice by $\eps$, so that we are discussing surfaces in
$\eps\Z^3$, the height function is defined as $1/\eps$ times
the coordinate sum, so that it is still integer-valued.
The normalized height function is $\eps$ times the height
function, and is the function which, when scaled by $\sqrt{3}$,
has graph which is the surface in $\eps\Z^3$.

Let $u:\partial U\to\R$ be a continuous function with the property that
$u$ can be extended to a Lipschitz function 
$\tilde u$ on the interior of $U$ having
the property
that the normal to the graph of 
$\tilde u$ has nonnegative coordinates, that is, the normal
points into the positive orthant $\R_{\ge0}^3$.
In other words, the graph of $u$ is a wire frame $\gamma$ of the type
discussed before.

Let $\bh\colon U\to\R$ be the {\bf asymptotic height function}  
with boundary values $u$, from Theorem \ref{ckpthm}. It is smooth
assuming the hypothesis of Theorem \ref{main1}.

Let $\nu=(p_a,p_b,p_c)$ be the normal vector to the graph of $\bh$,
scaled so that $p_a+p_b+p_c=1$. The directional derivatives of
$\bar h$ in the directions $\hat x,\hat y,\hat z$ are respectively
\begin{equation}\label{3p-1}
3p_c-1,3p_a-1,3p_b-1.
\end{equation}

\subsection{Measures and gauge equivalence}

We let $\mu=\mu(\G)$ be the uniform measure on 
dimer configurations on a finite
graph $\G$. 

If edges of $\G$ are given positive real weights, we can define a new
probability measure, the {\bf Boltzmann measure}, giving a configuration a 
probability proportional to the product of its edge weights. 

Certain edge-weight functions lead to the same Boltzmann measure: in particular
if we multiply by a constant the weights of all the edges in $\G$ having a
fixed vertex, the Boltzmann measure does not change,
since exactly one of these weights is used in every configuration.
More generally, two weight functions $\nu_1,\nu_2$ are said to be
{\bf gauge equivalent} if $\nu_1/\nu_2$ is a product of such
operations, that is, if there are functions $F_1$ on white
vertices and $F_2$ on black vertices so that for each edge $\w\b$,
$\nu_1(\w\b)/\nu_2(\w\b)=F_1(\w)F_2(\b).$ Gauge equivalent weights
define the same Boltzmann measure.

It is not hard to show that for 
planar graphs, two edge-weight functions are gauge equivalent
if and only if they have the same {\bf face weights}, where the weight
of a face is defined to be the alternating product of the edge weights
around the face (that is, the first, divided by the second, 
times the third, and so on), see e.g. \cite{KOS}.

In this paper we will only consider weights which are gauge 
equivalent to constant
weights (or nearly so), so the Boltzmann measure 
will always be (nearly) the uniform measure.

\subsection{Kasteleyn matrices}

Kasteleyn showed that one can count dimer configurations on planar graph 
with 
the determinant of the certain matrix, the ``Kasteleyn matrix"
\cite{Kast}.
In the current case, when the underlying graph is part of the honeycomb
graph, the Kasteleyn matrix $K$ is just the adjacency matrix from white vertices
to black vertices.

For more general bipartite planar graphs, and when the edges have
weights, the matrix is a signed, weighted
version of the adjacency matrix \cite{Percus}, whose determinant
is the sum of the weights of dimer coverings.
Each entry $K(\w,\b)$ is
a complex number with modulus given by the corresponding edge weight (or zero if
the vertices are not adjacent), and an argument which must be chosen in such a way
that around each face the alternating product of the entries
(the first,
divided by the second, times the third, and so on) is positive if the face
has $2\bmod 4$ edges and negative if the face has $0\bmod 4$ edges
(since we are assuming the graph is bipartite, each face has an even number of
edges. For nonbipartite graphs, a more complicated condition is necessary).

The Kasteleyn matrix is unique up to gauge transformations, which consist
of pre- and post-multiplication by diagonal matrices (with, in general,
complex entries). If the weights are real then we can choose a gauge
in which $K$ is real, although in certain cases it is convenient to allow 
complex numbers (we will below).

Probabilities of individual edges occurring in a random tiling can
likewise be computed using the minors of the inverse Kasteleyn matrix:
\begin{thm}[\cite{Kenyon.localstats}]\label{Kinvlocalstats}
The probability of edges $\{(\b_1,\w_1),\dots,(\b_k,\w_k)\}$ occurring
in a random dimer covering is 
$$\left(\prod_{i=1}^k K(\w_i,\b_i)\right)\det \{K^{-1}(\b_i,\w_j)\}_{1\leq i,j\leq k}.$$
\end{thm}

On an infinite graph $K$ is defined similarly
but $K^{-1}$ is not unique in general.
This is related to the fact that there are potentially many different measures
which could be obtained as limits of Boltzmann measures on 
sequences of finite graphs filling out the infinite graph.
The edge probabilities for these measures can all be described as in the theorem
above,
but where the matrix ``$K^{-1}$" now depends on the measure; see the next
section for examples.

\subsection{Measures in infinite volume}

On the infinite honeycomb graph $\H$ there is a two-parameter
family of natural translation-invariant
and ergodic probability measures on dimer configurations,
which restrict to the uniform measure on finite regions
(i.e. when conditioned on the complement of the finite region:
we say they are {\bf conditionally uniform}). 
Such measures are also known as 
{\bf ergodic Gibbs measures}.
They are classified in the following theorem due to Sheffield.

\begin{thm}[\cite{Sheff}]\label{Sheffthm}
For each $\nu=(p_a,p_b,p_c)$ with $p_a,p_b,p_c\geq0$ and scaled so 
that $p_a+p_b+p_c=1$  there is a
unique translation-invariant ergodic Gibbs measure $\mu_{\nu}$ on 
the set of dimer coverings of $\H$, for which the 
height function has average normal $\nu$. 
This measure can be obtained as the
limit as $n\to\infty$ 
of the uniform measure on the set of those dimer coverings of $\H_n=\H/nL$ whose proportion
of dimers in the three orientations is $(p_a:p_b:p_c)$, up to errors tending
to zero as $n\to\infty$.
Moreover every ergodic Gibbs measure
on $\H$ is of the above type for some $\nu$.
\end{thm}

The unicity in the above statement is a deep and important result.

Associated to $\mu_\nu$ is an infinite matrix, the {\bf inverse Kasteleyn matrix} 
of $\mu_\nu$,
$K^{-1}_{\nu}=(K^{-1}_\nu(\b,\w))$ whose rows index the black vertices and 
columns index the white vertices, and 
whose minors give local statistics for $\mu_{\nu}$, 
just as in Theorem \ref{Kinvlocalstats}.
From \cite{KOS} there is an explicit formula for $K_\nu^{-1}$: let
$\w=m_1\hat x+n_1\hat y$ and 
$\b=e_1+m_2\hat x+n_2\hat y=\w+e_1+m\hat x+n\hat y$ where
$m=m_2-m_1,n=n_2-n_1$.
Then
\begin{equation}\label{Knu_abc}
K^{-1}_{\nu}(\b,\w)= a(\frac{a}{b})^m(\frac{b}{c})^nK^{-1}_{abc}(\b,\w),
\end{equation}
where $a,b,c,z,w$ are as defined in section (\ref{abcdef}) and
\begin{eqnarray}
\label{Kinvabcdef}
K^{-1}_{abc}(\b,\w)&=&\frac1{(2\pi i)^2}\int_{|z|=|w|=1}\frac{z_1^{-m+n}w_1^{-n}}{a+bz_1+cw_1}\frac{dz_1}{z_1}\frac{dw_1}{w_1}\\
&=&
\label{Kinvabcasymp}
\frac1{\pi}\Im\left(\frac{z^{-m+n}w^{-n}}{cwm+an}\right)+
O\left(\frac{1}{m^2+n^2}\right).
\end{eqnarray}

This formula for $K_{abc}^{-1}$ and its asymptotics were derived in \cite{KOS}:
they are obtained from the limit $n\to\infty$ of the inverse 
Kasteleyn matrix on the torus $\H/nL$
with edge weights $a,b,c$ according to direction. It is not hard to check
from (\ref{Kinvabcdef}) that $KK^{-1}=Id$

From (\ref{Knu_abc}), the matrix $K_\nu^{-1}$ is just a 
gauge transformation of $K_{abc}^{-1}$,
that is, obtained by pre- and post-composing with diagonal matrices.

Defining $F(\w)=(bz/a)^{m_1}(cw/bz)^{n_1}$ and  
$F(\b)=a(bz/a)^{-m_2}(cw/bz)^{-n_2}$
we can write, using (\ref{Kinvabcasymp}),
\begin{equation}\label{Knu_F}
K^{-1}_{\nu}(\b,\w)=\frac1{\pi}\Im\left(\frac{F(\w)F(\b)}{cwm+an}
\right)+|F(\b)F(\w)|O(\frac{1}{m^2+n^2}).
\end{equation}
We'll use this function $F$ below.

As a sample calculation, the $\mu_\nu$-probability of a single horizontal edge, 
from $\w=0$ to $\b=e_1$, being present in a random dimer covering is 
(see Theorem \ref{Kinvlocalstats})
$$K_\nu^{-1}(\b,\w)=aK_{abc}^{-1}(\b,\w)=
\frac{a}{(2\pi i)^2}\int_{\T^2}\frac{1}{a+bz_1+cw_1}\frac{dz_1}{z_1}\frac{dw_1}{w_1}=
\frac{\theta_a}{\pi},$$
where $\theta_a$ is, as before, the angle opposite side $a$ in a 
triangle with sides $a,b,c$. This is consistent with (\ref{3p-1}).

\subsection{$T$-graphs}\label{Tgraphsection}

\subsubsection{Definition}

\begin{figure}[htbp]
\begin{center}\scalebox{.5}{\includegraphics{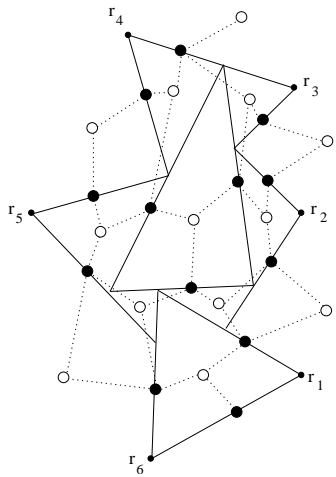}}
\end{center}
\caption{A T-graph (solid) and associated graph $\Gd$ (dotted).
\label{constTgraph}}
\end{figure}

$T$-graphs were defined and studied in \cite{KS}. A  pairwise disjoint
collection $L_1, L_2, \ldots, L_n$ of open line segments in $\R^2$
{\bf forms a T-graph in $\R^2$} if $\cup_{i=1}^n L_i$ is connected
and contains all of its limit points except for some finite set $R =
\{r_1, \ldots, r_m \}$, where each $r_i$ lies on the boundary of
the infinite component of $\R^2$ minus the closure of
$\cup_{i=1}^n L_i$. See Figure \ref{constTgraph} for an example
where the outer boundary is a polygon. Elements in $R$ are
called {\bf root vertices} and are labeled in cyclic order; 
the $L_i$ are called {\bf complete
edges}. We only consider the case that the outer boundary 
of the $T$-graph is a simple
polygon, and the root vertices are the convex corners of this polygon.
(An example where the outer boundary is not a polygon is a ``T''
formed from two edges, one ending in the interior of the other.)

\begin{figure}[htbp]
\begin{center}\scalebox{.35}{\includegraphics{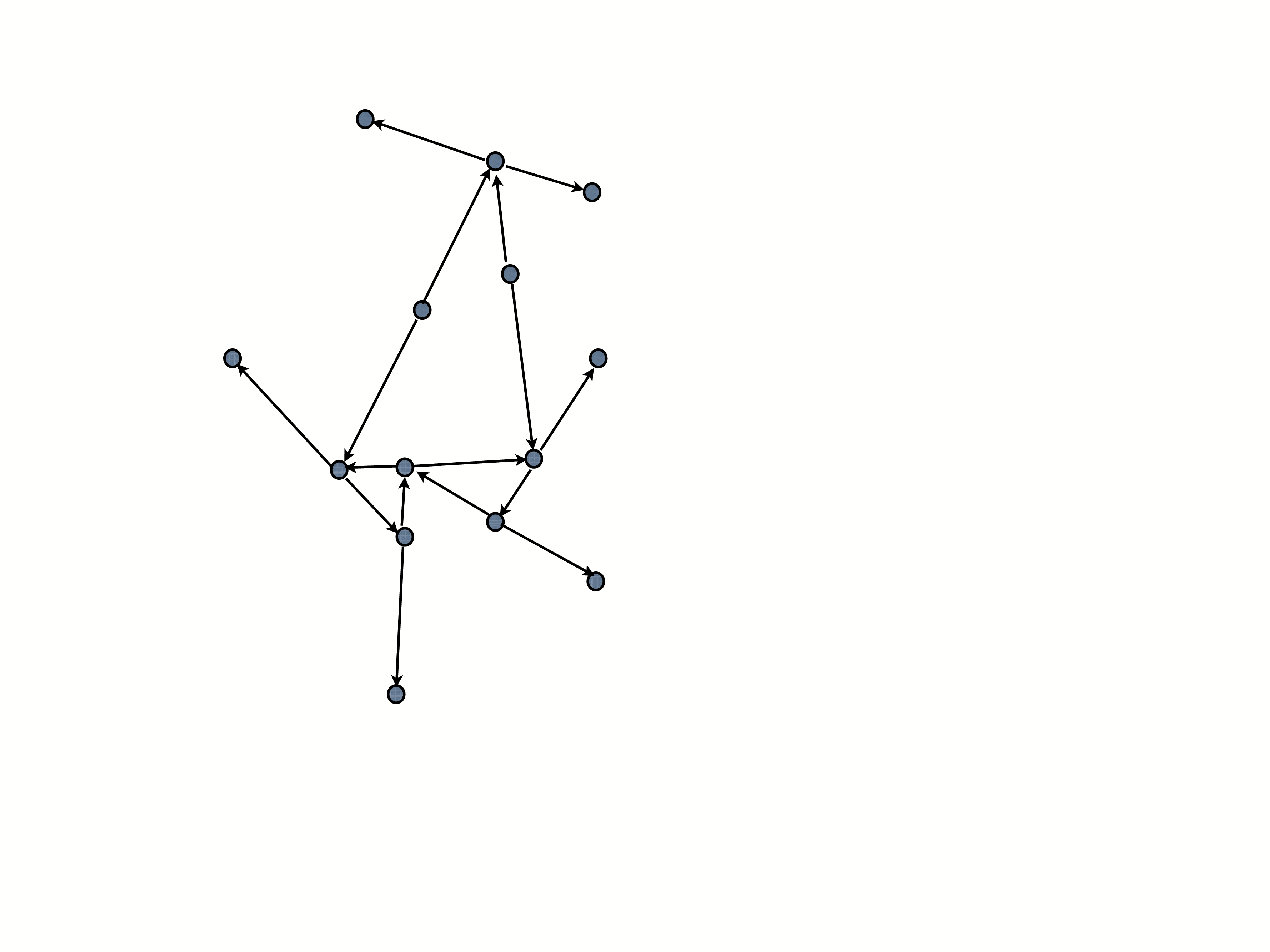}}\end{center}
\caption{The Markov chain associated to the $T$-graph of Figure
\ref{constTgraph}. Note that root vertices are sinks of the Markov chain.
\label{TgraphMC}}
\end{figure}

Associated to a $T$-graph is a Markov chain $\Gt$, 
whose vertices are the points which are
endpoints of some $L_i$. Each non-root vertex is in the interior
of a unique $L_j$ (because the $L_j$ are disjoint); 
there is a transition from that vertex to its
adjacent vertices along $L_j$, and the 
transition probabilities are proportional to the inverses of the Euclidean
distances. Root vertices are sinks of the 
Markov chain. 
See Figure \ref{TgraphMC}.

Note that the coordinate functions on $\Gt$
are harmonic functions on $\Gt\setminus R$. 
More generally, any
function $f$ on $\Gt$ which is harmonic on $\Gt\setminus R$
(we refer to such functions as harmonic functions on $\Gt$)
has the property that it is linear along
edges, that is, if $v_1,v_2,v_3$ are vertices on the same complete
edge then
\begin{equation}\label{tlinear}
\frac{f(v_1)-f(v_2)}{v_1-v_2}=\frac{f(v_2)-f(v_3)}{v_2-v_3}.
\end{equation}

\subsubsection{Associated dimer graph and Kasteleyn matrix}\label{Tgraph}

Associated to a $T$-graph is a weighted bipartite planar graph
$\Gd$ constructed as follows, see Figure \ref{constTgraph}. 
Black vertices of $\Gd$ are the $L_j$. 
White vertices
are the bounded complementary regions, as well as one white vertex  
for each boundary path joining
consecutive root vertices $r_j$ and $r_{j+1}$ (but not for the path 
from $r_m$ to $r_1$).
The complementary regions are called {\bf faces}; the paths between 
adjacent root vertices
are called {\bf outer faces}.
Edges connect the $L_i$ to each
face it borders along a positive-length subsegment. 
The edge weights are equal to the
Euclidean length of the bounding segment. 

To $\Gd$ there is a canonically
associated Kasteleyn matrix of $\Gd$: this is the $n\times
n$ matrix $K_{\Gd}=(K_{\Gd}(\w,\b))$ with rows indexing the white vertices and
columns indexing the black vertices of $\Gd$. We have $K_{\Gd}(\w,\b)=0$
if $\w$ and $\b$ are not adjacent, and otherwise $K_{\Gd}(\w,\b)$ is the
complex number equal to the edge vector corresponding to the edge
of the region $\w$ along complete edge $\b$ (taken in 
the counterclockwise direction around $\w$). In particular $|K_{\Gd}(\w,\b)|$ is the
length of the corresponding edge of $\w$. 

\begin{lemma}
$K_{\Gd}$ is a Kasteleyn matrix for $\Gd$, that is, the alternating product 
of the matrix entries for edges around a bounded face is positive real or 
negative real according
to whether the face has $2\bmod 4$ or $0\bmod 4$ edges,
respectively.
\end{lemma}

By alternating product we mean the first, divided by the second, times the third,
etc. 

\begin{figure}[htbp]
\begin{center}\scalebox{.8}{\includegraphics{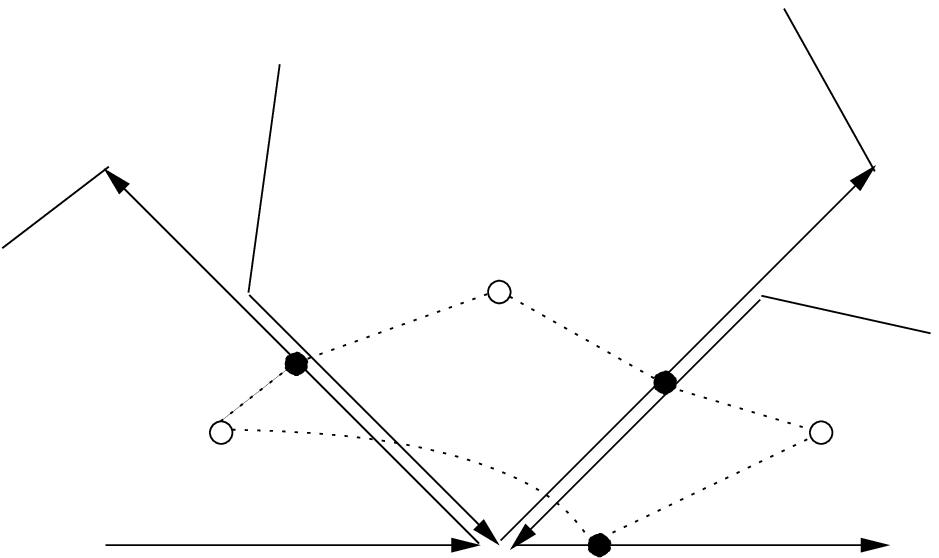}}\end{center}
\caption{\label{face}A face of $\Gd$ (dotted).}
\end{figure}

\begin{proof}
Let $f$ be a bounded face of $\Gd$ (we mean not one of the outer faces). 
It corresponds to a meeting point of two or more
complete edges; this meeting point is in the interior of exactly 
one of these complete edges, $L$.  See Figure \ref{face}. 
In $\Gd$, for each other black vertex on that face
the two edges of the $T$-graph to neighboring white vertices
have opposite orientations. The two edges parallel to $L$ 
(horizontal in the figure) have the same orientation, so their ratio is positive.
This implies the result. 
\end{proof}

Although we won't need this fact, in \cite{KS} it is shown that the set
in-directed spanning forests of $\Gt$ (rooted at the root vertices
and weighted by the product of the transition probabilities)
is in measure-preserving (up to a global constant)
bijection with the set of dimer coverings of $\Gd$.

\subsubsection{Harmonic functions and discrete analytic functions}
\label{harm/anal}

To a harmonic function $f$ on a $T$-graph $\Gt$ we associate a
derivative $df$ which is a function on black vertices of $\Gd$ as follows.
Let $v_1$ and $v_2$ be two distinct points on complete edge
$\b$, considered as complex numbers. We define 
\begin{equation}\label{df}
df(\b)=\frac{f(v_2)-f(v_1)}{v_2-v_1}.
\end{equation}
Since $f$ is linear along any complete edge (equation (\ref{tlinear})), 
$df$ is independent of the choice
of $v_1$ and $v_2$.

\begin{lemma}\label{daflemma}
If $f$ is harmonic on a $T$-graph $\Gt$ and $K=K_{\Gd}$ is the associated
Kasteleny matrix, then $\sum_{\b\in B}K(\w,\b)df(\b)=0$ 
for any interior white vertex $\w$.
\end{lemma}

\begin{proof}
Let $\b_1,\dots,\b_k$ be the neighbors of $\w$ in cyclic order. To
each neighbor $\b_i$ is associated a segment of a complete edge
$L_i$. Let $v_i$ and $v_{i+1}$ be the endpoints of that segment,
and $w_{i}$ and $w_i'$ be the endpoints of $L_i$. Then
$$\frac{f(v_{i+1})-f(v_i)}{v_{i+1}-v_i}=\frac{f(w_i')-f(w_i)}{w_i'-w_i}$$
since the harmonic function is linear along $L_i$. In particular
$$K_{\Gd}(\w,\b_i)df(\b_i)=(v_{i+1}-v_i)\frac{f(w_i')-f(w_i)}{w_i'-w_i}=
f(v_{i+1})-f(v_i).$$
Summing over $i$ (with cyclic indices) yields the result.
\end{proof}

Note that at a boundary white vertex $\w$, 
$\sum_BK_{\Gd}(\w,\b)df(\b)$ is the difference
in $f$-values at the adjacent root vertices of $\Gt$.

We will refer to a function $g$ on black vertices of $\Gd$ 
satisfying $\sum_{\b\in B}K_{\Gd}(\w,\b)g(\b)=0$
for all interior white vertices $\w$ as a {\bf discrete analytic function}.

The construction in the above lemma can be reversed, starting from a
discrete analytic function $df$ (on black vertices of $\Gd$) and integrating to get
a harmonic function $f$ on $\Gt$: define $f$ arbitrarily at a vertex of $\Gt$
and then extend to neighboring vertices (on a same complete edge)
using (\ref{df}). The extension is well-defined by discrete analyticity.

\subsubsection{Green's function and $K_{\Gd}^{-1}$}\label{CS.green}\label{CS.Kinv}

We can relate $K_{\Gd}^{-1}$ to the conjugate Green's function on $\Gt$ 
using the construction of the previous section, as
follows. 

\begin{figure}
\begin{center}\scalebox{.4}{\includegraphics{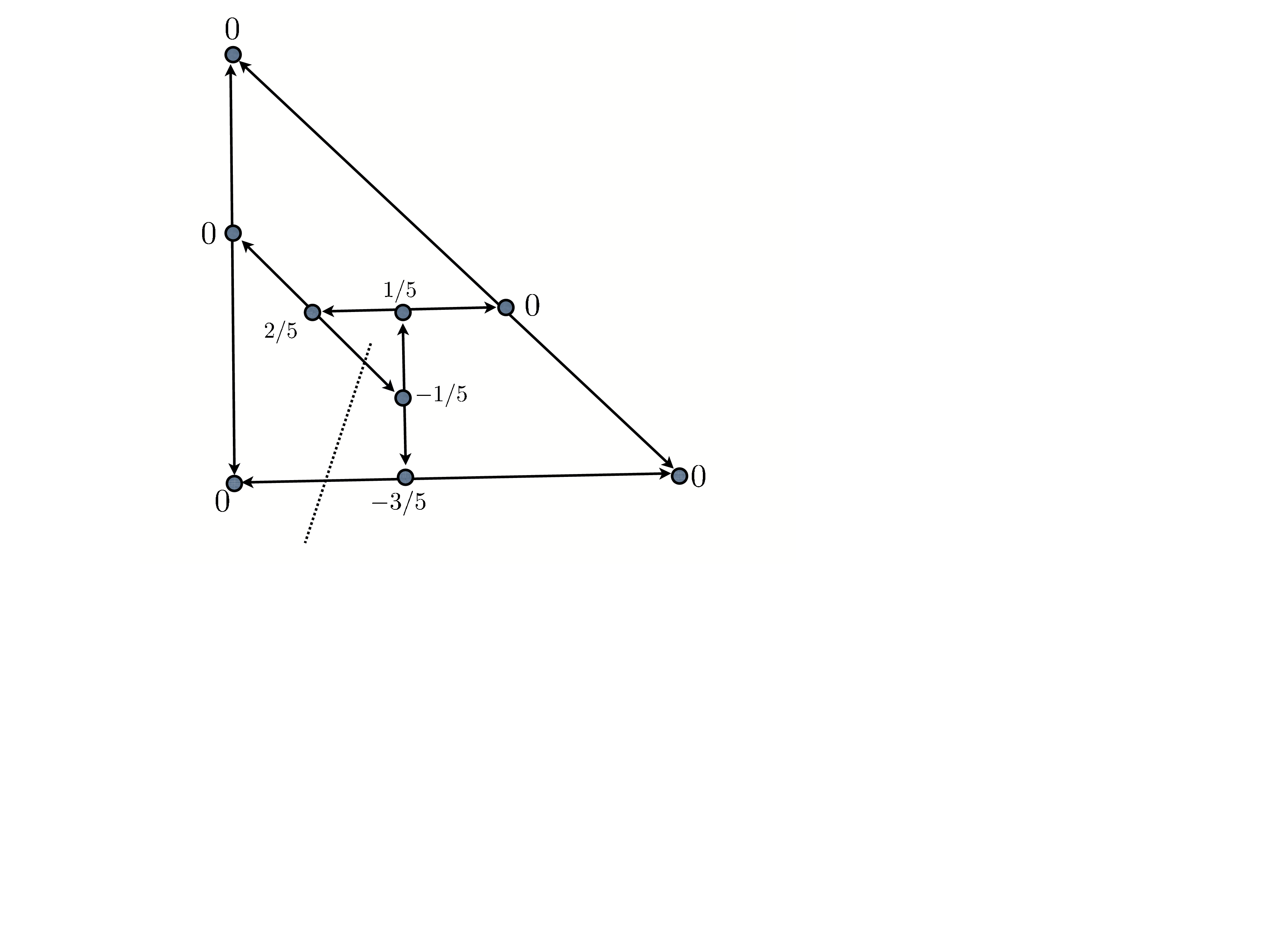}}
\end{center}
\caption{Conjugate Green's function example. Here the sides of the triangle
are bisected in ratio $2:3$ and interior complete edges $1:1$.\label{Gconj}}
\end{figure}

Let  $\w$ be an interior
face of $\Gt$, and $\ell$ a path from a point in $\w$ to the outer boundary of $\Gt$
which misses all the vertices of $\Gt$. 
For vertices $v$ of $\Gt$, define the {\bf conjugate
Green's function} $G^*(\w,v)$ to be the expected algebraic number of crossings
of $\ell$ by the random walk started at $v$ and stopped at the boundary.
This is the unique function with zero boundary values which is harmonic 
everywhere except for a jump discontinuity of $-1$ across $\ell$ when going 
counterclockwise around $\w$.
(If there were two such functions, their difference would be harmonic everywhere
with zero boundary values. ) See Figure \ref{Gconj} for an example.

Let $K_{\Gd}^{-1}(\b,\w)$ be the discrete analytic function of $\b$ defined from
$G^*(\w,v)$ as in the previous section; 
on an edge which crosses $\ell$, 
define $K_{\Gd}^{-1}(\b,\w)$
using two points on $\b$ on the same side of $\ell$.  This function clearly
satisfies $K_{\Gd} K_{\Gd}^{-1}=I$ by Lemma \ref{daflemma}, and therefore
is independent of the choice of $\ell$.

Note that $K_{\Gd}^{-1}$ also has the following probabilistic interpretation: take two particles, started simultaneously 
at two different points $v_1,v_2$ of the same complete edge, and couple their
random walks so that they start independently, take simultaneous steps, and 
when they meet they stick together for all future times.
Then the difference in their winding numbers around $\w$ is determined
by their crossings of $\ell$ before they meet. 
That is, $K_{\Gd}^{-1}(\b,\w)(v_1-v_2)$
is the expected difference in crossings before the particles meet or until
they hit the boundary, whichever comes first. 

\section{Constant-slope case}\label{constslope}

In this section we compute the asymptotic expansion
of $K^{-1}_{\Gd}$ (Theorem \ref{constcase}) in
the special case is when $\gamma$ is planar. In this case the normalized 
asymptotic height function $\bh$ is linear, and its normal $\nu$ is 
constant. This case already contains 
most of the complexity of the general case, which is treated in
section \ref{gencase}. 

Let
$\nu=(p_a,p_b,p_c)$ be the normal to $\bh$, scaled as usual 
so that $p_a+p_b+p_c=1$. 
The angles $\theta_a,\theta_b,\theta_c$
are constant, and we choose $a,b,c$ as before to be constant as
well. Define a function 
$$
F(\w)=(bz/a)^m(cw/bz)^n
$$ 
at a white vertex
$\w=m\hat x+n\hat y$ and 
$$F(\b)=a(bz/a)^{-m}(cw/bz)^{-n}$$ at a black
vertex $\b=e_1+m\hat x+n\hat y$. 
These functions are defined on all of the honeycomb graph $\H$.
Let $K_\H$ be the adjacency matrix of $\H$ (which, as we mentioned earlier,
is a Kasteleyn matrix for $\H$).

\begin{lemma}\label{closedlemma} We have 
\begin{equation}\label{closed}
\sum_{\b} K_\H(\w,\b)F(\b)=0=\sum_{\w}F(\w)K_\H(\w,\b)
\end{equation}
where $K_\H$ is the 
adjacency matrix of $\H$. 
\end{lemma}

\begin{proof} This follows from the equation $a+bz+cw=0$.
\end{proof}

\subsection{$T$-graph construction}\label{GT.CS}

Define a $1$-form $\Omega$ on edges of $\H$ by
\begin{equation}\label{constgauge}
\Omega(\w\b)=-\Omega(\b\w)=2\Re(F(\w))F(\b).
\end{equation} 

By (\ref{closed}) 
the dual form $\Omega^*$ (defined by $\Omega^*((\w\b)^*)=\Omega(\w\b)$) is
closed (the integral around any closed cycle is zero) 
and therefore $\Omega^*=d\Psi$ for a complex-valued 
function $\Psi$ on
$\H^*$. Here $\H^*$, the dual of the honeycomb, is the graph of the equilateral
triangulation of the plane. Extend $\Psi$ linearly over the edges of $\H^*$. 
This defines a mapping from $\H^*$ to $\C$ with the
property that the images of the white faces are
triangles similar to the $a,b,c$-triangle (via
orientation-preserving similarities), and the images of black
faces are segments. This follows immediately from the definitions:
if $\b_1,\b_2,\b_3$ are the three neighbors of a white vertex $\w$ of $\H$,
the edges $\w\b_i$ have values
$2\Re(F(\w))F(\b_i)$ which are proportional to $F(\b_1):F(\b_2):F(\b_3)$,
which in turn are proportional to $a:bz:cw$ by (\ref{closed}).
If $\w_1,\w_2,\w_3$ are the three neighbors of a black vertex then
the corresponding edge values are $2\Re(F(\w_i))F(\b)$
which are proportional to $\Re(F(\w_1)):\Re(F(\w_2)):\Re(F(\w_3))$,
that is, they all have the same slope and sum to zero.

\begin{figure}
\begin{center}\scalebox{1}{\includegraphics{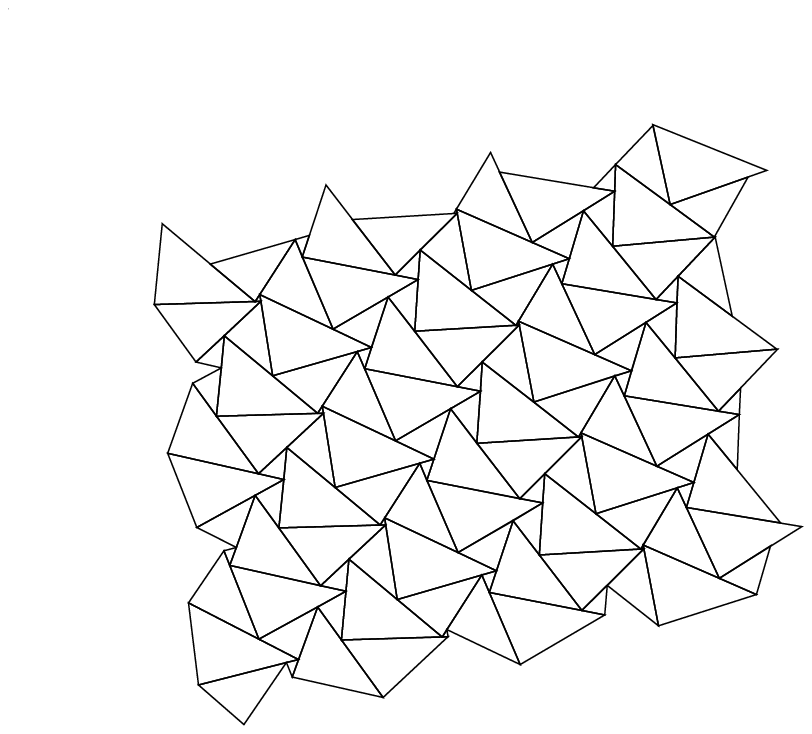}}
\end{center}
\caption{\label{bigconstTgraph}The $\Psi$-image of $\H^*$ is a $T$-graph
covering $\R^2$}
\end{figure}

It is not hard to see that the images of the white triangles are in fact 
non-overlapping (see \cite{KS}, Section 5 for the proof, or look at Figure \ref{bigconstTgraph}).
It may be that $\Re F(\w)=0$ for some $\w$; in this case choose a 
generic modulus-$1$ complex number $\lambda$ and replace $F(\w)$
by $\lambda F(\w)$ and $F(\b)$ by $\overline{\lambda}F(\b)$. So we can 
assume that each white triangle is similar to the $a,b,c$-triangle.
In fact, this operation will be important later; note that 
by varying $\lambda$ the size of an individual triangle varies;
by an appropriate choice we can make any particular triangle have maximal
size (side lengths $a,b,c$).

\begin{lemma}
The mapping $\Psi$ is almost linear, that is, it is a linear map
$\phi(m,n)=cwm+an$
plus a bounded function. 
\end{lemma}

\begin{proof}
Consider for example a vertical column
of horizontal edges $\{\w_1\b_1,\dots,\w_k\b_k\}$ of $\H$
connecting a face $f_1$ to face $f_{k+1}$. 
We have 

\begin{eqnarray}\label{polybdy}
\sum_{i=1}^k \Omega^*((\w_i\b_i)^*)&=&\sum_{i=1}^k \Omega(\w_i\b_i)\notag \\
&=&\sum_{i=1}^k 
( F(\w_i)+\overline{F(\w_i)})F(\b_i) \notag\\
&=&ka+\sum_{i=1}^k\overline{F(\w_i)}F(\b_i)\\\notag
&=&ka+a\sum_{i=1}^k z^{2i}w^{-2i}\\\notag
&=&\phi(f_{k+1})-\phi(f_1)+osc,
\end{eqnarray}
where $osc$ is oscillating and in fact 
$O(1)$ independently of $k$ (by hypothesis $z/w\not\in\R$). In the other two lattice directions
the linear part of $\Psi$ is again $\phi$, so that $\Psi$ is almost
linear in all directions.
\end{proof}

This lemma shows that the image of $\H^*$ under $\Psi$ is an infinite
$T$-graph $\Ht$ covering all of $\R^2$. The
images of the black triangles are the complete edges and have lengths $O(1)$.

If we insert a $\lambda$ as above and let $\lambda$ vary over the unit circle, 
one sees all possible local structures of the 
$T$-graph, that is, the geometry of the $T$-graph $\Ht=\Ht(\lambda)$ 
in a neighborhood of a triangle
$\Psi(\w)$ only depends up to homothety
on the argument of $\lambda F(\w)$.
\label{lambdasection}

Recall that $\G^*$ is a subgraph of $\eps\H^*$ approximating $U$. 
We can restrict $\Psi$ to $\frac1\eps\G^*$ thought of as a subgraph of $\frac1\eps\H$, and then multiply
its image by $\eps$. 
Thus we get a finite sub-$T$-graph $\Gt$ of $\eps\Ht$. Let 
$\Psi_{\eps}=\eps\circ\Phi\circ\frac{1}{\eps}$ so that $\Psi_{\eps}$ acts on $\G^*$.

The union of
the $\Psi_\eps$-images of the white triangles in $\G^*$ forms a polygon
$P$. Define the ``dimer'' graphs $\H_D$ associated to $\Ht,$ and $\Gd$ associated to
$\Gt$ as 
in section \ref{Tgraph}.  See
Figure \ref{constTgraph} for the $T$-graph arising from the graph
$\G^*$ of Figure \ref{G*}. Note that $\Gd$ contains $\G$
(defined in section \ref{dualgraph})
but has extra white vertices along the boundaries. These extra vertices
make the height function on $\Gd$ approximate the desired linear function 
(whose graph has normal $\nu$), see the next section.

From (\ref{constgauge}) we have
\begin{lemma}
Edge weights of $\H_D$ and $\Gd$ are gauge equivalent to constant edge weights.
Indeed, for $\w$ not on the boundary of $\Gd$ we have 
\begin{equation}
\label{gaugeequivKGD}
K_{\Gd}(\w,\b)=2\eps\Re( F(\w))F(\b)K_\G(\w,\b)
\end{equation}
and similarly for all $\w$,
\begin{equation}\label{gaugeequivKHD}
K_{\H_D}=2\Re(F(\w))F(\b)K_\H(\w,\b).
\end{equation}
\end{lemma}

Asymptotics of $K_{\Gd}^{-1}$ are described below in section \ref{constKinv}.
For $K_{\H_D}$, from (\ref{Knu_F}) and (\ref{gaugeequivKHD}) we have
\begin{equation}\label{Kinvplane}
K_{\H_D}^{-1}(\b,\w)=\frac{1}{2\pi\Re( F(\w))F(\b)}\Im\left(
\frac{ F(\w)F(\b)}{\phi(\b)-\phi(\w)}\right)+O(\frac{1}{|\phi(\b)-\phi(\w)|^2}).
\end{equation}
Here $K^{-1}_{\H_D}$ is the inverse constructed from the conjugate
Green's function on the $T$-graph $\H_T$. We use the fact that
$2\Re(F(\w))F(\b)K^{-1}_{\H_D}(\b,\w)$ coincides with $K^{-1}_{\nu}$
of (\ref{Knu_F}) since both satisfy the equation that $dK^{-1}$ 
equals the conjugate Green's function.

\subsection{Boundary behavior}\label{boundarybehavior}

Recall that from our region $U$ we constructed a graph $\G$
(section \ref{dualgraph}). From the normalized height function
$u$ on $\partial U$ (which is the restriction of a linear function to $\partial U$)
we constructed the $T$-graph $\Gt$ and then the dimer graph $\Gd$. 

In this section we show that the normalized boundary height function of $\Gd$
when $\Gd$ is chosen as above
approximates $\bh$. This is proved in a roundabout way: we first show that
the boundary height does not depend (up to local fluctuations)
on the exact choice of boundary conditions, as long as we construct $\Gt$
from $\G^*$ as in section \ref{graphs}. Then we compute the height change
for ``simple" boundary conditions.

\subsubsection{Flows and dimer configurations}

Any dimer configuration $m$ defines a flow (or $1$-form) $[m]$
with divergence $1$ at each white vertex
and divergence $-1$ at each black vertex: just flow by $1$ along 
each edge in $m$ and $0$ on the other
edges. 

The set $\Omega_1\subset[0,1]^E$ of unit white-to-black flows 
(i.e. flows with divergence $1$ at each white vertex and $-1$
at each black vertex) and with capacity
$1$ on each edge is a convex polytope whose vertices are
the dimer configurations \cite{LP}. On the graph
$\H$ define the flow $\omega_{1/3}\in\Omega_1$
to be the flow with value $1/3$ on each edge $\w\b$ from $\w$ to $\b$. 
Up to a factor $1/3$, this flow can be used to define the height
function, in the sense that for any dimer configuration $m$,
$[m]-\omega_{1/3}$ is a divergence-free flow
and the integral of its dual (which is closed) is $1/3$ times the height
function of $m$. 
That is, the height difference between two points
is three times the flux of $[m]-\omega_{1/3}$
between those points. This is easy to see: across an edge which
contains a dimer the height changes by
$\pm2$ (depending on the orientation); if the edge does not contain a dimer the height changes by $\mp1$, for so that the height change can be written
$\pm(3\id-1)$ where $\id$ is the indicator function of a dimer on that edge.

For a finite subgraph of $\H$ the {\it boundary} height 
function can be obtained by integrating around the boundary
($3$ times) the dual of $[m]-\omega_{1/3}$ for some $m$.

\subsubsection{Canonical flow}\label{canflow}

\begin{figure}[htbp]
\begin{center}\scalebox{.5}{\includegraphics{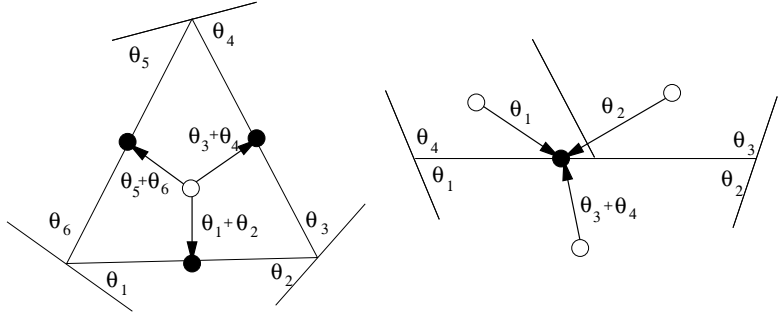}}
\end{center}
\caption{\label{flowb}Defining the canonical flow (divide 
the angles by $2\pi$).}
\end{figure}

On $\H_D$ there is a canonical flow $\omega\in\Omega_1$
defined as illustrated in Figure
\ref{flowb}: let $v_1v_2$ be the two vertices of $\Gt$
along $\b$ adjacent to $\w$. The flow from $\w$ to $\b$ is $1/(2\pi)$ times
the sum of the two angles
that the complete edges through $v_1$ and $v_2$ make with $\b$.
One or both of these angles may be zero.

Note that the total flow out of $\w$ is $1$.
For an edge $\b$, the total flow into $\b$ is also $1$ as illustrated 
on the right in Figure
\ref{flowb}.

Now $\Gd$ is a subgraph of $\H_D$ with extra white
vertices around its boundary. The canonical flow on $\H_D$
restricts to a flow on $\Gd$ except on edges connecting to 
these extra white vertices (we define the canonical flow there to be zero). 
This flow has divergence $1,-1$ at
white/black vertices, except at the black vertices of $\Gd$
connected to the boundary white vertices, and the boundary white vertices themselves.
If $m$ is a dimer covering of $\Gd$, the flow $[m]-\omega$
is now a divergence-free flow on $\Gd$ except at 
these black and white boundary vertices.

\begin{lemma}
Along the boundary of $\Gd$
the divergence of $[m]-\omega$ for any dimer configuration
$m$ is the turning angle of the boundary of $P$.
\end{lemma}

\begin{proof}
Consider a complete edge $L$ corresponding to black vertex $\b$.
The canonical flow into $\b$ has a contribution from the 
two endpoints of $L$. The flow $[m]$ can be considered to contribute $-1/2$
for each endpoint. 

Recall that each endpoint of $L$ ends in the interior of another
complete edge or at a convex vertex of the polygon $P$.
If an endpoint of $L$ ends in the interior of another complete edge,
and there are white faces adjacent to the two sides of this endpoint
(that is, the endpoint is not a concave vertex of $P$)
then the contribution of the canonical flow is also $-1/2$, so the contribution
of $[m]-\omega$ is zero. 

Suppose the endpoint is 
at a concave vertex $v$ of $P$ with exterior angle
$\theta<\pi$. 
The contribution
from $v$ to the flow of $[m]-\omega$ into $L$ is $-1/2+\frac{\theta}{2\pi}$.
The other complete edge at $v$ does not end at $v$ and so has no contribution.
This quantity is $1/2\pi$ times the turning angle of the boundary.

Suppose the endpoint is at a convex vertex $v'$ of $P$ of interior angle $\theta<\pi$. There is some complete
edge $L'$ of $\Ht$, containing $v'$ in its interior, which is not in $\Gt$.
The sum of the contributions of $[m]-\omega$ for the endpoints of the two
complete edges $L_1,L_2$ of $\Gt$ meeting at $v'$
is $-1/2-\theta/2\pi$. 
This is the $1/2\pi$ times the turning angle at the convex vertex,
minus $1$. 

The contribution from $[m]$ from the boundary white vertices is
$1$ per white boundary vertex, that is, one per convex vertex of $P$.
\end{proof}

This lemma proves in particular that the divergence of $[m]-\omega$ 
is bounded for any $[m]$.

\subsubsection{Boundary height}
Recall that the height function of a dimer covering $m$ can be defined
as the flux of $\omega_{1/3}-[m]$. 
In particular, the flux of $\omega_{1/3}-\omega$ defines the boundary height function
up to $O(1)$ (the turning of the boundary),
since the flux of $[m]-\omega$ is the boundary turning angle. 

The flux of $\omega_{1/3}-\omega$ between two faces can be computed along 
any path, and in fact because both $\omega_{1/3}$ and $\omega$ are 
locally defined from $\H_D$, we see that the
flux does not depend on the choice of the nearby boundary.
Let us compute this flux and show that it is linear along lattice directions,
and therefore linear everywhere in $\H_D$.

\begin{figure}[htbp]
\begin{center}\scalebox{1}{\includegraphics{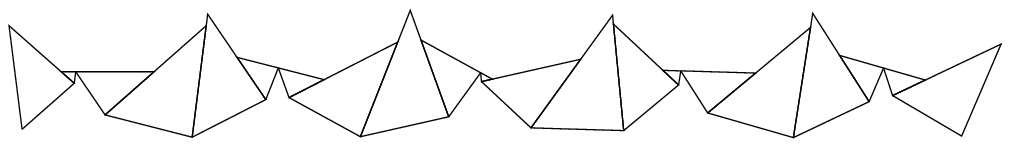}}
\end{center}
\caption{\label{Gdbdy1}A column of triangles from $\Gt$.}
\end{figure}

Take a vertical column of
horizontal edges in $\G$, and let us compute $\omega$ on this set of edges.
The $\Psi$ image of the dual of this column is a polygonal
curve $\eta$ whose $j$th edge is (using (\ref{polybdy})) 
a constant times $1+(\frac{z}{w})^{2j}$.
The image of the triangles in the vertical column of $\G^*$ 
is as shown in Figure \ref{Gdbdy1}.

The $j$th edge is part of a complete edge corresponding to the $j$th black vertex
in the column. By the argument of the previous section,
the flux is equal to the number of convex corners of this polygonal curve $\eta$,
that is, corners where the curve turns left.
The curve $\eta$ has a convex corner when 
$$\Im\left(\frac{1+(z/w)^{2j+2}}{1+(z/w)^{2j}}\right)\geq 0,$$
that is, using $z/w=-e^{i\theta_a},$ when 
$2j\theta_a\in [\pi,\pi+2\theta_a].$
Assuming that $\theta_a$ is irrational, this happens with frequency
$\frac{2\theta_a}{2\pi}$,
so the flux of $\omega_{1/3}-\omega$ along a column of length $n$ is
$n(\frac{\theta_a}{\pi}-\frac13)$, and the average flux per edge is $p_a-1/3$.
Therefore the average height change per horizontal edge is $3p_a-1$.
If $\theta_a$ is rational, a continuity argument shows that $3p_a-1$ is
still the average height change per horizontal edge.
A similar result holds in the other directions, and (\ref{3p-1})
show that the average height has normal $(p_a,p_b,p_c)$ as desired.

\subsection{Continuous and discrete harmonic functions}\label{contdisc}

To understand the asymptotic expansion of $G^*$, the conjugate
Green's function on $\Gt$, from which we can get $K^{-1}$, we need two ingredients.
We need to understand
the conjugate Green's function on $\Ht$, and also 
the relation between continuous harmonic functions
on domains in $\R^2$ and discrete harmonic functions on (domains in) $\eps\Ht$.

\subsubsection{Discrete and continuous Green's functions}\label{greenderivation}

On $\H_T$, 
the discrete conjugate Green's function $G^*(\w,v)$, for $\w\in\H$ and $v\in\Ht$,
can be obtained 
from integrating the exact formula for $K^{-1}_\nu$ given in (\ref{Knu_abc})
as discussed in Section \ref{CS.green}.

As we shall see, this formula differs even in its leading term from 
the continuous conjugate Green's
function, due to the singularity at the diagonal. Basically, because our 
Markov chain is directed, a long random walk can have a nonzero expected
winding number around the origin. This causes the conjugate Green's function,
which measures this winding,
to have a component of $\Re\log v$ as well as a part $\Im\log v$.

The continuous conjugate Green's function on the whole plane is 
$$g^*(v_1,v_2)=\frac1{2\pi}\Im\log(v_2-v_1).$$
(We use $G^*$ to denote the discrete conjugate Green's function
and $g^*$ the continuous version.)

To compute the discrete conjugate Green's function $G^*$,
if we simply restrict $g^*$ to the vertices of $\eps\Ht$,
it will be very nearly harmonic as a function of $v_2$ 
for large $|v_2-v_1|$, but the discrete Laplacian of $g^*$ at
vertices $v_2$ near $v_1$ 
(within $O(\eps)$ of $v_1$) will be of constant order in general. 
We can correct for the non-harmonicity at a vertex $v'$ by adding
an appropriate multiple of the actual (non-conjugate) discrete
Green's function $G(v,v_2)$. The large scale behavior of
this correction term is a constant times $\Re\log(v_2-v').$ 
So we can expect the long-range behavior of the discrete conjugate 
Green's function $G^*(\w,v)$, for $|\w-v|$ large, 
to be equal to $g^*(\w,v)$ plus a sum of terms
involving the real part $\Re\log(v-v')$ for $v'$s within $O(\eps)$ of $\phi(\w)$.
These extra terms sum to a function of the form 
$c\log|v-\phi(\w)|+\eps s(v)+O(\eps^2),$ where $s$ is a smooth function and $c$ is a
real constant, both $c$ and $s$ depending on the local 
structure of $\Ht$ near $\phi(\w)$.

This form of $G^*$ can be seen explicitly, of course, if we integrate
the exact formula (\ref{Kinvplane}). We have
\begin{lemma}\label{Gplane}
The discrete conjugate Green's function $G^*$ on 
the plane $\eps\Ht$ is asymptotically
\begin{multline}\label{Gplaneformula}
G^*(\w,v)=\frac1{2\pi}\left(\Im\log(v-\phi(\w))+
\frac{\Im( F(\w))}{\Re( F(\w))}
\Re\log(v-\phi(\w))\right)+\\
+\eps s(\w,v)+O(\eps^2)+O(\frac1{|v-\phi(\w)|})
\end{multline}
\begin{equation}\label{gstarasymp}
=\frac1{2\pi\Re( F(\w))}
\Im\bigl( F(\w)\log(v-\phi(\w))\bigr)+\eps s(\w,v)
+O(\eps^2)+O(\frac1{|v-\phi(\w)|})
\end{equation}
where $s(\w,v)$ is a smooth function of $v$.
\end{lemma}
\begin{proof} From the argument of the previous paragraph,
it suffices to compute the constant in front of the 
$\Re\log(v-\phi(\w))$ term. 
This can be computed by differentiating the above formula 
with respect to $v$ and comparing with formula (\ref{Kinvplane}) for $K^{-1}$.
The differential of $\eps s(\w,v)$ is $O(\eps)$.
Let $v_1,v_2$ be two vertices of complete edge $\b$,
coming from adjacent faces of $\H$, adjacent across an edge $\b\w_1$
(so that $v_1-v_2=\Omega^*(\b\w_1)=2\eps\Re( F(\w_1))F(\b)$).
We have for $\w$ far from $\w_1$
\begin{eqnarray*}
K_{\H_D}^{-1}(\b,\w)&=&\frac{G^*(\w,v_1)-G^*(\w,v_2)}{v_1-v_2}\\
&=&
\frac{\frac1{2\pi}\Im\left(\frac{ F(\w)}{
\Re( F(\w))}\frac{v_1-v_2}{\phi(\b)-\phi(\w)}\right)}{
2\eps\Re( F(\w_1))F(\b)}+O(\eps)\\
&=&\frac1{2\pi\Re( F(\w))F(\b)}
\Im\left(\frac{ F(\w)F(\b)}{\phi(\b)-\phi(\w)}\right)
+O(\eps).
\end{eqnarray*}
\end{proof}

Note that in fact for any complex number $\lambda$ of modulus $1$,
we get a discrete conjugate Green's function $G^*_\lambda$ on the 
graph $\eps\Ht(\lambda)$ (from Section \ref{lambdasection})
with similar asymptotics.

\subsubsection{Smooth functions}

Constant and linear functions on $\R^2$, when restricted
to $\eps\Ht$, are exactly harmonic. 
More generally, if we take a continuous harmonic function $f$ on $\R^2$ and 
evaluate it on the vertices of $\eps\Ht$, the result will be close to 
a discrete harmonic function $f_\eps$, in the sense that the
discrete Laplacian will be $O(\eps^2)$: if $v$ is a vertex of $\eps\Ht$
and $v_1,v_2$ are its (forward) neighbors located at $v_1=v-\eps d_1e^{i\theta}$ and
$v_2=v+\eps d_2e^{i\theta}$ then the Taylor expansion of $f$ about $v$ yields
$$\Delta f(v)=f(v)-\frac{d_2}{d_1+d_2}f(v-\eps d_1e^{i\theta})-
\frac{d_1}{d_1+d_2}f(v+\eps d_2e^{i\theta})=
O(\eps^2).$$ 
This situation is not as good as in the (more standard) case of a graph like
$\eps\Z^2$, where if we evaluate a continuous harmonic function on the vertices,
the Laplacian of the resulting discrete function is $O(\eps^4)$:
$$f(v)-\frac14\Big(f(v+\eps)+f(v+\eps i)+f(v-\eps)+f(v-\eps i)\Big)=O(\eps^4).$$ 

In the present case the principal error is due to the second derivatives of $f$.
To get an error smaller than $O(\eps^2)$,
we need to add to $f$ a term which cancels out the  $\eps^2$ error.
We can add to $f$ a function which is 
$\eps^{2}$ times a bounded function $f_2$
whose value at a point $v$ depends only on the local structure
of $\eps\Ht$ near $v$ and on the second derivatives of $f$ at $v$.

\begin{lemma} For a smooth harmonic function $f$ on $\R^2$ whose
second partial derivatives don't all vanish at any point, there is a 
bounded function $f_\eps$ on $\Ht$ such that
$$f_\eps(z)=f(z)+\eps^2 f_2(z)$$
has discrete Laplacian of order $O(\eps^3)$, 
where $f_2(z)$ depends only the second derivatives of $f$ at $z$ 
and on the local structure of the graph $\Ht$ at $z$.
\end{lemma}

\begin{proof}
We have exact formulas for one discrete harmonic function, 
the conjugate Green's function on $\eps\H_T$, 
and we know its asymptotics (Lemma \ref{Gplane}), equation (\ref{gstarasymp}), 
which are 
$$G^*(\w,z_2)\approx\eta(z_1,z_2)=\frac1{2\pi}\Im(c\log(z_2-z_1))$$ for a constant $c$ depending on 
the local structure of the graph near $\w$, and 
where $z_1$ is a point in $\phi(\w)$. We'll let
$\w$ be the origin in $\eps\H_T$ and $z_1=0$; $\eta(0,z_2)$
is a continuous harmonic function of $z_2$.

For $z\in\R^2$ consider the second derivatives of the function $f$,
which by hypothesis are not all zero.
There is a point $z_2=\beta(z)\in\R^2$ at which $\eta(0,z_2)$
has the same second derivatives.
Indeed, $f$ has three second partial derivatives, $f_{xx},f_{xy},$ and $f_{yy}$,
but because $f$ is harmonic $f_{yy}=-f_{xx}$. We have 
$\frac{\partial}{\partial z_2}\eta(0,z_2)=\frac1{2pi}\Im\frac{c}{z_2}$,
and the second partial derivatives of $\eta$ are the real and imaginary parts
of $const/z_2^2$, which is surjective, in fact $2$ to $1$,
as a mapping of $\R^2-\{0\}$ to itself. 
In particular there are two choices of $z_2$
for which $f(z)$ and $\eta(\beta(z))$ have the same second derivatives.
Since $f$ and $\eta$ are smooth, by taking a consistent choice,
$\beta$ can be chosen to be a smooth function as well.

Consider the function 
$$f_\eps(z)=f(z)+G_\lambda^*(0,\beta(z))-\eta(0,\beta(z)),$$
where $G_\lambda^*$ is the discrete conjugate Green's function
and $\lambda$ is chosen so that $\Ht(\lambda)$ (see section \ref{lambdasection})
has local structure at $\beta(z)$
identical to that of $\Ht$ at $z$.

We claim that the discrete Laplacian of $f_\eps$ is $O(\eps^3)$.
This is because $G^*_\lambda(0,\beta(z))$ is discrete harmonic,
and $f(z)-\eta(0,\beta(z))$ has vanishing second derivatives.

We also have that $G^*(0,\beta(z))-\eta(0,\beta(z))=O(\eps^2)$, see Lemma 
\ref{Gplane} above. 
\end{proof}

\subsection{$K_{\Gd}^{-1}$ in constant-slope case}\label{constKinv}

\begin{thm}\label{constcase}
In the case of constant slope $\nu$ and a bounded domain $U$, 
let $\xi$ be a conformal diffeomorphism from
$\phi(U)$ to $\Hy$.
When $\b$ and $\w$ are converging to different points as $\eps\to0$
we have 
\begin{equation}\label{consteqn}
K_{\Gd}^{-1}(\b,\w)=
\frac{1}{2\pi\Re( F(\w))F(\b)}\Im\left(\frac{\xi'(\phi(\b))F(\w)
F(\b)}{\xi(\phi(\b))-\xi(\phi(\w))}+
\frac{\xi'(\phi(\b))
\overline{F(\w)}F(\b)}{\xi(\phi(\b))-\overline{\xi(\phi(\w))}}\right)+O(\eps).
\end{equation}
\end{thm}

\begin{proof}
The function $G^*_{\Gt}(\w,v)$ is equal to the function (\ref{Gplaneformula}) for the whole
plane, plus a harmonic function on $\Gt$ whose boundary values are
the negative of the values of (\ref{Gplaneformula}) on the boundary of $\Gt$.

Since discrete harmonic functions on $\Gt$ are close to continuous
harmonic functions on $U$, we can work with the corresponding
continuous functions.

From (\ref{gstarasymp}) we have 
\begin{equation}\label{Gagain}
G^*_{\Ht}(\w,v)=\frac1{2\pi}\Im \left(\frac{F(\w)}{\Re(F(\w))}\log(v-z_1)\right)+\eps s(z_1,v)+O(\eps)^2,
\end{equation}
where $z_1$ is a point in face $\phi(\w)$.
The continuous harmonic function of $v$ on $U$ whose values on $\partial U$
are the negative of the values of (\ref{Gagain}) on $\partial U$ 
is
\begin{multline}\label{these2}
-\frac1{2\pi}\Im\left( \frac{F(\w)}{\Re F(\w)}\log(v-z_1)\right)+\\
\frac1{2\pi}\Im\left(\frac{F(\w)}{\Re F(\w)}\log(\xi(v)-\xi(z_1))+
\frac{\overline{F(\w)}}{\Re F(\w)}\log(\xi(v)-\overline{\xi(z_1)})\right)+\eps s_2(z_1,v)+O(\eps)^2,
\end{multline}
where $s_2$ is smooth.

The discrete Green's function $G^*_{\Gt}$ must be the sum of 
(\ref{Gagain}) and (\ref{these2}):
$$G^*_{\Gt}(\w,v)= \frac1{2\pi}\Im\left(\frac{F(\w)}{\Re F(\w)}\log(\xi(v)-\xi(z_1))+
\frac{\overline{F(\w)}}{\Re F(\w)}\log(\xi(v)-\overline{\xi(z_1)})\right)+
\eps s_3(z_1,v)+O(\eps)^2.$$

Differentiating gives the result (as in Lemma \ref{Gplane}).
\end{proof}

It is instructive to 
compare the discrete conjugate Green's function in the above proof with 
the continuous conjugate Green's function on $U$ which is $$g^*(z_1,z_2)=\frac1{2\pi}\Im\log(\xi(z_1)-\xi(z_2))+\frac1{2\pi}
\Im\log(\xi(z_1)-\overline{\xi(z_2)}).$$

\subsubsection{Values near the diagonal}

Note that when $\b$ is within $O(\eps)$ of $\w$, 
and neither is close to the boundary,
the discrete Green's function $G^*(\w,v)$ for $v$ on $\b$
is equal to the discrete Green's
function on the plane $G_{\Ht}^*(\w,v)$ plus an error which
is $O(1)$ coming from the corrective term due to the boundary. The error is
smooth plus oscillations of order $O(\eps^2)$, so that within
$O(\eps)$ of $\w$ the error is
a linear function plus $O(\eps)$. 
Therefore when we take derivatives
$$K^{-1}_{\Gd}(\b,\w)=K^{-1}_{\H_D}(\b,\w)+O(1),$$ 
which, since $K^{-1}_{\H_D}$ is of order $O(\eps^{-1})$ when $|\b-\w|=O(\eps)$,
implies that the local statistics are given by $\mu_{\nu}$. 
\begin{thm}\label{localstatconst}
In the case of constant slope $\nu$, 
the local statistics at any point in the interior of $U$ are
given in the limit $\eps\to0$ 
by $\mu_{\nu}$, the ergodic Gibbs measure on tilings
of the plane of slope $\nu$.
\end{thm}

\section{General boundary conditions}\label{gencase}

Here we consider the general setting: $U$ is a smooth Jordan domain and
$\bh$, the normalized asymptotic height function on $U$, is not necessarily
linear.

\subsection{The complex height function}\label{gaugeheight}

The equation (\ref{pde}) implies that the form
$$\log(\Phi-1)d\x-\log(\frac1{\Phi}-1)d\y$$
is closed. 
Since $U$ is simply connected it is $dH$ for a 
function $H\colon U\to\C$ which we call the {\bf complex height function}.

The imaginary part of $H$ is related to $\bh$: 
we have $\arg(\Phi-1)=\pi-\theta_c$ (Figure \ref{triangles})
and $\arg(\frac{1}{\Phi}-1)=\theta_a-\pi$,
which gives
$\Im \,dH=(\pi-\theta_c)d\x+(\pi-\theta_a)d\y.$
{}From (\ref{3p-1}) we have $d\bh=(3p_c-1)d\x+(3p_a-1)d\y,$
so $$\frac3{\pi}\Im \,dH=2(d\x+d\y)-d\bh.$$ The
real part of $H$ is the logarithm of a special gauge function which
we describe below.

We have
\begin{eqnarray}
H_{\x} &=& \log(\Phi-1)\\
H_{\y} &=& -\log(\frac1{\Phi}-1)\\
H_{\x\x} &=& \frac{\Phi_{\x}}{\Phi-1}=\frac{-\Phi\Phi_{\y}}{\Phi-1}\\
H_{\y\y} &=&- \frac{1}{\Phi-1}\frac{\Phi_{\y}}{\Phi}.
\end{eqnarray}

\subsection{Gauge transformation}

The mapping $\Phi$ is a real analytic mapping from $U$ to the
upper half plane. It is an open mapping
since $\Im(\Phi_{\x}/\Phi_{\y})=-\Im\Phi\neq0$, but may have isolated critical 
points. The Ahlfors-Bers theorem gives us a 
diffeomorphism $\phi$ from $U$ onto the upper half plane
satisfying the Beltrami equation
$$\frac{\frac{d\phi}{d\bar z}}{\frac{d\phi}{dz}}=
\frac{\frac{d\Phi}{d\bar z}}{\frac{d\Phi}{dz}}=\frac{\Phi-e^{i\pi/3}}{\Phi-e^{-i\pi/3}},$$ that is
$\phi_{\x}=-\Phi \phi_{\y}$. Such a $\phi$ exists by the Ahlfors-Bers
theorem \cite{AB}. It follows that $\Phi$ is of the form $f(\phi)$ for
some holomorphic function $f$ from $\Hy$ into $\Hy$.
Since $\partial U$ is smooth, $\phi$ is smooth up to and 
including the boundary, and $\phi_{\x},\phi_{\y}$ are both nonzero.

For white vertices of $\G$ define
\begin{equation}\label{Fdef}
F(\w)=e^{\frac1{\eps}H(\w)}\sqrt{\phi_{\y}(\w)}(1+\eps M(\w)),
\end{equation}
where
$M(\w)$ is any function which satisfies
\begin{multline}\label{Meqn}
e^{-H_{\x}}M_{\x}-e^{H_{\y}}M_{\y}=e^{-H_{\x}}\left(\frac{H_{\x\x}^2}8-\frac{H_{\x\x\x}}6-
\frac{H_{\x\x}\phi_{\x\y}}{4\phi_{\y}}-\frac{\phi_{\x\y}^2}{8\phi_{\y}^2}+
\frac{\phi_{\x\y\y}}{4\phi_{\y}}\right)+\\
e^{H_{\y}}\left(\frac{H_{\y\y}^2}8+
\frac{H_{\y\y\y}}6+\frac{H_{\y\y}\phi_{\y\y}}{4\phi_{\y}}-\frac{\phi_{\y\y}^2}{8\phi_{\y}^2}+
\frac{\phi_{\y\y\y}}{4\phi_{\y}}\right).
\end{multline}

The existence of such an $M$ follows from the fact that
the ratio of the coefficients of $M_{\x}$ and $M_{\y}$ is $-e^{H_{\x}+H_{\y}}=\Phi$,
so (\ref{Meqn}) is of the form 
$$M_{\x}+\Phi M_{\y}=J(x,y)$$
for some smooth function $J$. This is the $\bar\partial$ equation
in coordinate $\phi$. We don't need to know $M$ explicitly; the final
result is independent of $M$. We just need its existence to get better estimates
on the error terms in Lemma \ref{calc} below.

For black vertices $\b$ define 
\begin{equation}\label{Fdef2}
F(\b)=e^{-\frac1{\eps}H(\w)}\sqrt{\phi_{\y}(\w)}(1-\eps M(\w)),
\end{equation}
where $\w$ is the vertex adjacent to and left of $\b$ and 
$M(\w)$ is as above.

\begin{lemma}\label{calc} For each black vertex $\b$ with 
three neighbors in $\G$ we have
$$\sum_{\w}F(\w)K(\w,\b)=O(\eps^3)$$ and for each white vertex $\w$ we have
$$\sum_{\b}K(\w,\b)F(\b)=O(\eps^3).$$
\end{lemma}

\begin{proof}
This is a calculation.
Let $\w,\w-\eps \hat x,\w+\eps \hat y$ be the three neighbors of $\b$.
Then, setting $H=H(\w)$, $\phi_{\y}=\phi_{\y}(\w)$, and $M=M(\w)$ we have
$$F(\w-\eps \hat x)=e^{\frac1{\eps}(H-\eps H_{\x}+\frac{\eps^2}2H_{\x\x}
-\frac{\eps^3}{6}H_{\x\x\x}+O(\eps^4))}
\sqrt{\phi_{\y}-\eps\phi_{\y\x}+\frac{\eps^2}{2}\phi_{yxx}+O(\eps^3)}
(1+\eps M-\eps^2 M_{\x}+O(\eps^3))$$
$$F(\w+\eps \hat y)=e^{\frac1{\eps}(H+\eps H_{\y}+\frac{\eps^2}2H_{\y\y}
+\frac{\eps^3}{6}H_{\y\y\y}+O(\eps^4))}
\sqrt{\phi_{\y}+\eps\phi_{\y\y}+\frac{\eps^2}{2}\phi_{\y\y\y}+O(\eps^3)}
(1+\eps M+\eps^2 M_{\y}+O(\eps^3)).$$
The sum of the leading order terms in $F(w)+F(w-\eps\hat x)+F(w+\eps \hat y)$ is
$$e^{\frac1{\eps}H}\sqrt{\phi_{\y}}
(1+e^{-H_{\x}}+e^{H_{\y}})=e^{\frac1{\eps}H}\sqrt{\phi}(1+\frac1{\Phi-1}+
\frac{\Phi}{1-\Phi})=0.$$
The sum of the terms of order $\eps$ is $\frac{\eps}2
e^{\frac1{\eps}H}\sqrt{\phi_{\y}}$ times
$$e^{-H_{\x}}(-\frac{\phi_{\y\x}}{\phi_{\y}}+H_{\x\x})+
e^{H_{\y}}(\frac{\phi_{\y\y}}{\phi_{\y}}+H_{\y\y})=$$
$$=\frac1{\Phi-1}\left(\frac{\Phi\phi_{\y\y}}{\phi_{\y}}+\Phi_{\y}+\frac{-\Phi\Phi_{\y}}{\Phi-1}\right)+
\frac{\Phi}{1-\Phi}\left(\frac{\phi_{\y\y}}{\phi_{\y}}-\frac{\Phi_{\y}}{\Phi(\Phi-1)}\right)=0$$
and the sum of the order-$\eps^2$ terms is
$\eps^2 e^{\frac1{\eps}H}\sqrt{\phi_{\y}}$ times
\begin{multline}
-e^{-H_{\x}}M_{\x}+e^{H_{\y}}M_{\y}+e^{-H_{\x}}\left(\frac{H_{\x\x}^2}8-\frac{H_{\x\x\x}}6-
\frac{H_{\x\x}\phi_{\x\y}}{4\phi_{\y}}-\frac{\phi_{\x\y}^2}{8\phi_{\y}^2}+
\frac{\phi_{\x\y\y}}{4\phi_{\y}}\right)+\\
e^{H_{\y}}\left(\frac{H_{\y\y}^2}8+
\frac{H_{\y\y\y}}6+\frac{H_{\y\y}\phi_{\y\y}}{4\phi_{\y}}-\frac{\phi_{\y\y}^2}{8\phi_{\y}^2}+
\frac{\phi_{\y\y\y}}{4\phi_{\y}}\right)=0.
\end{multline}

A similar calculation holds at a white vertex, and we get the same expression
for the $\eps^2$ contribution (changing the signs of $M, H, d/d\x$ and $d/d\y$ 
gives the same expression).
\end{proof}

It is clear from this proof that the error in the statement
can be improved to any order $O(\eps^{3+k})$
by replacing $M(\w)$ with $M_0(\w)+\eps M_1(\w)+
\dots+\eps^k M_k(\w)$, where each $M_j$ satisfies an equation of
the form (\ref{Meqn}) except with a different right hand side---the 
right-hand side will depend on derivatives
of $H,\phi$ and the $M_i$ for $i<j$. 
For our proof below we need an error $O(\eps^4)$ and therefore
the $M_0$ and $M_1$ terms, even though the final
result will depend on neither $M_0$ nor $M_1$.

\subsection{Embedding}\label{Gdconstruction}\label{gen.CS}

Define a $1$-form 
$$\Omega(\w\b)=2\Re(F(\w))F(\b)K(\w,\b)$$
on 
edges of $\G$, where $F$ is defined in (\ref{Fdef},\ref{Fdef2}) 
(since $K(\w,\b)=\eps$ is a constant,
this is an unimportant factor for now,
but in a moment we will perturb $K$). 
By the comments after the proof of Lemma \ref{calc},
the dual form $\Omega^*$ on $\G^*$ can be chosen to be
closed up to $O(\eps^4)$ and so
there is a function $\tilde\phi$ on $\G^*$, defined up to an additive constant, 
satisfying $d\tilde\phi=\Omega^*+O(\eps^3)$.

In fact up to the choice of the additive constant, $\tilde\phi$ is equal to 
$\phi$ plus an oscillating function.
This can be seen as follows.
For a horizontal edge $\w\b$ we have 
$F(\w)F(\b)=\phi_{\y}(\w)+O(\eps^2).$
Thus on a vertical column $\{\w_1\b_1,\dots,\w_k\b_k\}$ of horizontal edges we
have 
$$\sum_{i=1}^k \Omega^*(\w_i\b_i)=\eps\sum_{i=1}^k 
( F(\w_i)+\overline{ F(\w_i)})F(\b_i) 
=\eps\sum\phi_{\y}(\w_i)+
\eps\sum_{i=1}^k\overline{F(\w_i)}F(\b_i)+O(\eps^3).$$
The first sum gives the change in $\phi$ from one endpoint of 
the column to the other, 
and the second sum is oscillating ($\overline{F(\w_i)}$ 
and $F(\b_i)$ have the same argument
which is in $(0,\pi)$ and which is a continuous function of the position)
and so contributes $O(\eps)$. Similarly, in the other two lattice directions
the sum is given by the change in $\phi$ plus an oscillating term.

Therefore by choosing the additive constant appropriately,
$\tilde\phi$ maps $\G^*$ to a small neighborhood of $\Hy$ 
in the spherical metric on $\hat\C$,
which shrinks to $\Hy$ as $\eps\to0$. 

The map $\tilde\phi$ has the following additional properties.
The image of the three edges of a black face of $\G^*$ are nearly
collinear, and the image of a white triangle is a triangle nearly similar
to the $a,b,c$-triangle, and of the same orientation. Thus it is nearly
a mapping onto a $T$-graph. In fact near a point where the relative weights
are $a,b,c$ (weights which are slowly varying on the scale of the lattice) the 
map is up to small errors
the map $\Psi$ of section \ref{constslope}. 

We can adjust the
mapping $\tilde\phi$ by $O(\eps^3)$ so that
the image of each black triangle is an exact line
segment: this can be arranged by choosing for each black face a
line such that the $\tilde\phi$-image of the corresponding black face 
is within $O(\eps^3)$ of that line;
the intersections of these lines can then be used to define a new mapping
$\Psi\colon\G^*\to\C$ which is an exact $T$-graph mapping. 

The mapping $\Psi$ will then correspond to the above $1$-form $\Omega$
but for a matrix $\tilde K$ with slightly 
different edge weights. Let us check how much the weights
differ from the original weights (which are $\eps$). 
As long as the triangular faces are
of size of order $\eps$ (which they are typically), 
the adjustment will change the edge lengths locally by $O(\eps^3)$
and therefore their relative lengths by $1+O(\eps^2)$. 
There will be some isolated triangular faces, however, which will 
be smaller---of order $O(\eps^2)$
because of the possibility that $ F(\w)$ might be 
nearly pure imaginary.
We can deal with these as follows. 
Once we have readjusted the ``large" triangular
faces we have an exact $T$-graph mapping on most of the graph.
We can then multiply $F(\w)$ by $\lambda=i$
and $F(\b)$ by $\overline{\lambda}=-i$: 
the readjusted weights give (for most of the graph)
a new exact $T$-graph mapping
(because we now have $\tilde KF=F\tilde K=0$ exactly for these weights),
but now all faces which were too small 
before become $O(\eps)$ in
size and we can readjust their dimensions locally by a factor $1+O(\eps^2)$.

In the end we have an exact mapping $\Psi$ of $\G^*$ onto a $T$-graph $\Gt$
and it distorts the edge weights of $K$ by at most $1+O(\eps^2)$. 
We shall see in section \ref{Kapprox} that this is close enough to get
a good approximation to $K^{-1}$. 

In conclusion the Kasteleyn matrix for $\Gd$ is equal
to $2\Re(F(\w))F(\b)\tilde K(\w,\b)$ where $\tilde K$ 
has edge weights $\eps+O(\eps^3)$.

\subsection{Boundary}

Along the boundary we claim that the normalized height function of $\Gd$
follows $u$. Since $\Gd$ arises from a $T$-graph, we can use
its canonical flow (section \ref{canflow}). Near any given point the canonical
flow looks like the canonical flow in the constant-weight
case---since the weights vary continuously, they vary slowly 
at the scale $\eps$, the scale of the graph. 
Since the canonical flow defines the slope of the normalized height function,
we have pointwise convergence of the derivative of $\bh$ along the boundary
to the derivative of $u$. Thus $\bh$ converges to $u$.
In fact this argument shows that the normalized height function of the canonical
flow converges to the asymptotic height function in the interior of $U$ as well.

\section{Continuity of $K^{-1}$}\label{Kapprox}

In this section we show how $K^{-1}$ changes under a small change in edge weights.

\begin{lemma}
Suppose $0<\delta\ll\eps$.
If $\G'$ is a graph identical to $\Gd$ but with
edge weights which differ by a factor $1+O(\delta)$, then
$$K^{-1}_{\Gd}=K^{-1}_{\G'}(1+O(\delta/\eps)).$$
\end{lemma}

In particular since $\delta=\eps^2$ in our case this will be
sufficient to approximate $K^{-1}$ to within $1+O(\eps)$.

\begin{proof}
For any matrix $A$ we have
$$(K+\delta\eps A)^{-1}=K^{-1}\left(1+\sum_{j=1}^\infty(-1)^j(\delta\eps)^j(AK^{-1})^j\right),$$
as long as this sum converges.

From Theorem \ref{Kinvasymp} below we have that $K_{\G'}^{-1}(\b,\w)=O(1/|\b-\w|)$.
When $A$ represents a bounded, weighted adjacency matrix of $\Gd$,
the matrix norm of $AK^{-1}$ is then
$$\|AK^{-1}\|\leq \max_{\w'}\sum_{\w}|\sum_{\b} A(\w',\b)K^{-1}(\b,\w)|\leq 3a\max_{\w'}\sum_{\w\neq\w'}\frac{1}{|\w'-\w|}=
O(\eps^{-2})$$
where $a$ is the maximum entry of $A$.
In particular when $\delta\ll\eps$ we have
$$(K+\delta\eps A)^{-1}=K^{-1}(1+O(\delta/\eps)).$$
\end{proof}

\section{Asymptotic coupling function}\label{gen.green}

We define $K_{\Gd}$ as in (\ref{gaugeequivKGD}) using the values 
(\ref{Fdef},\ref{Fdef2}) for $F$ (and $K_{\G}$ is the adjacency matrix of $\G$).

\begin{thm} \label{Kinvasymp} If $\b,\w$ are not within $o(1)$ of the boundary of $U$,
we have
\begin{equation}\label{Kinvas}K_{\Gd}^{-1}(\b,\w)=\frac{1}{2\pi \Re( F(\w))F(\b)}
\Im\left(\frac{ F(\w)F(\b)}{\phi(\b)-\phi(\w)}+
\frac{\overline{ F(\w)}F(\b)}{\phi(\b)-\overline{\phi(\w)}}\right)+O(\eps).
\end{equation}
When one or both of $\b,\w$ are near the boundary, but they are not
within $o(1)$ of each other, then $K_{\Gd}^{-1}(\b,\w)=O(1)$. 
\end{thm}

\begin{proof}
The proof is identical to the proof in the constant-slope case, see Theorem
\ref{constcase}, except that $\xi\phi$ there is $\phi$ here. The $\xi'$
factors from (\ref{consteqn}) are here absorbed in the definition of the functions 
$\phi$ and $F$.
\end{proof}

As in the case of constant slope, when $\b$ and $\w$ are close
to each other (within $O(\eps)$) and not within $o(1)$ of the boundary, 
Theorem 
\ref{localstatconst} applies to show that the local statistics
are give by $\mu_{\nu}$.

\section{Free field moments}

As before let $\phi$ be a diffeomorphism from $U$ to  $\Hy$ satisfying
$\phi_{\x}=-\Phi\phi_{\y}$ where $\Phi$ is defined from $\bh$ as in 
section \ref{gaugeheight}.
 
\begin{thm}\label{main}
Let $\bh$ be the asymptotic height function on $\Gd$, $h_\eps$ the 
normalized height function of a random tiling, 
and $\hat h=\frac{2\sqrt{\pi}}{3\eps}(h_\eps-\bh)$.
Then $\hat h$ converges weakly as $\eps\to0$ to $\phi^*{\cal F},$ the pull-back
under $\phi$ of ${\cal F}$, the Gaussian free field on $\Hy$. 
\end{thm}

Here weak convergence means that for any smooth test function $\psi$ on $U$,
zero on the boundary, we have 
$$\eps^2\sum_{\Gd}\psi(f)\hat h(f)\to \int_U\psi(x){\cal F}(\phi(x))|dx|^2,$$
where the sum on the left is over faces $f$ of $\Gd$.

\begin{proof}
We compute the moments of $\cal F$.
Let $\psi_1,\dots,\psi_k$ be smooth functions on $U$,
each zero on the boundary. 
We have
$$\E\left[\left(\eps^2\sum_{f_1\in\Gd}\psi_1(f_1)\hat h(f_1)\right)\cdots
\left(\eps^2\sum_{f_k\in\Gd}\psi_k(f_k)\hat h(f_k)\right)\right]=
$$
$$=\eps^{2k}\sum_{f_1,\dots,f_k}\psi_1(f_1)\cdots\psi_k(f_k)
\E[\hat h(f_1)\dots\hat h(f_k)].$$
From Theorem \ref{mom} below the sum becomes
\old{$$\int_U\cdots\int_U
\prod \psi_i(x_i)|dx_i|^2\cdot\left[
\sum_{\text{pairings }\sigma}\prod_{j=1}^{k/2} 
G(\phi(s_{\sigma(2j-1)}),\phi(s_{\sigma(2j)}))\right]+o(1)$$}
$$=\int_U\cdots\int_U \E[{\cal F}(\phi(x_1))\dots{\cal F}(\phi(x_k))]
\prod \psi_i(x_i)|dx_i|^2+o(1),$$ 
that is, the moments of $\hat h$
converge to the moments of the free field $\phi^*({\cal F})$.
Since the free field is a Gaussian process, it is determined by its moments.
This completes the proof.
\end{proof}

\begin{thm}\label{mom}
Let $s_1,\dots,s_k\in U$ be distinct points in the interior of $U$.
For each $\eps$ let $f_1,f_2,\dots,f_{k}$ be faces of $\Gd$, 
with $f_i$ converging to $s_i$ as $\eps\to0$.
If $k$ is odd we have 
$$\lim_{\eps\to0}\E[\hat h(f_1)\dots\hat h(f_k)]
=0$$
and if $k$ is even we have
$$\lim_{\eps\to0}\E[\hat h(f_1)\dots\hat h(f_k)]=
\sum_{\text{pairings }\sigma}\prod_{j=1}^{k/2} 
G(\phi(s_{\sigma(2j-1)}),\phi(s_{\sigma(2j)}))$$
where $$G(z,z')=-\frac1{2\pi}\log\left|\frac{z-z'}{z-\bar z'}\right|$$ 
is the Dirichlet Green's
function on $\Hy$ and the sum is over all pairings of the indices.

If two or more of the $s_i$ are equal, we have
$$\E[\hat h(f_1)\dots\hat h(f_k)]=O(\eps^{-\ell}) ,$$
where $\ell$ is the number of coincidences (i.e. $k-\ell$ is the number of 
distinct $s_i$).
\end{thm}

\begin{proof}
We first deal with the case that the $s_i$ are distinct.
Let $\gamma_1,\dots,\gamma_{k}$ be pairwise disjoint paths of faces from 
points $s_i'$ on the boundary to the $f_i$. We assume that these paths 
are far apart from each other (that is,
as $\eps\to0$ they converge to disjoint paths). 
The height $h_\eps(f_i)$ 
can be measured as a sum along $\gamma_i$. 

We suppose without loss 
of generality that each $\gamma_i$ is a polygonal path consisting 
of a bounded number of straight segments which are parallel
to the lattice directions $\hat x,\hat y,\hat z$. In this case, 
by additivity of the height change along $\gamma_i$
and linearity of the moment in each index, 
we may as well assume that $\gamma_i$ is in a single lattice direction. 

Now the change in $\hat h$ along $\gamma_i$ is given by
the sum of $a_{ij}-\E(a_{ij})$ where $a_{ij}$ is the indicator 
function of the $j$th edge 
crossing $\gamma_i$ (with a sign according the the direction of $\gamma_i$). 
So the moment is 
$$\E[\hat h(f_1)\dots\hat h(f_{k})]=
\sum_{j_1\in\gamma_1,
\dots,j_k\in\gamma_k}\E[(a_{1j_1}-\E[a_{1j_1}])\dots(a_{kj_k}-
\E[a_{kj_k}])].$$

If $\w_{ij_i},\b_{ij_i}$ are the vertices of edge $a_{ij_i}$, 
this moment becomes (see \cite{K.confinv})
{\tiny
$$\sum_{j_1,\dots,j_k}\left(\prod_{i=1}^k K(\w_{ij_i},\b_{ij_i})
\right)\left|\begin{array}{cccc}
0&K^{-1}(\b_{2j_2},\w_{1j_1})&\dots&K^{-1}(\b_{kj_k},\w_{1j_1})\\
K^{-1}(\b_{1j_1},\w_{2j_2})&0\\
\vdots&&\ddots&\vdots\\
K^{-1}(\b_{1j_1},\w_{kj_k})&\dots&K^{-1}(\b_{k-1j_{k-1}},\w_{kj_k})&0
\end{array}\right|.
$$}

\noindent In particular, the effect of subtracting off the mean values of the $a_{ij_i}$
is equivalent to cancelling the diagonal terms $K^{-1}(b_{ij_i},w_{ij_i})$
in the matrix.

Expand the determinant as a sum over the symmetric group.
For a given permutation $\sigma$, which must be fixed-point free
or else the term is zero,  
we expand out the corresponding
product, and sum along the paths. 
For example if $\sigma$ is the $k$-cycle $\sigma=(1 2 \dots k)$, the 
corresponding term is
$$\text{sgn}(\sigma)\sum_{j_1,\dots,j_k}\left(\prod_{i=1}^k K(w_{ij_i},b_{ij_i})\right) 
K^{-1}(\b_{1j_1},\w_{2j_2})K^{-1}(\b_{2j_2},\w_{3j_3})\cdots 
K^{-1}(\b_{kj_k},\w_{1j_1})=$$

{\tiny
\begin{multline}=-(\frac{-1}{4\pi i})^k\sum_{j_1,\dots,j_k} \left(
\frac{F(\b_{1j_1}) F(\w_{2j_2})}{\phi(\b_{1j_1})-\phi(\w_{2j_2})}+
\frac{F(\b_{1j_1})\overline{ F(\w_{2j_2})}}{\phi(\b_{1j_1})-
\overline{\phi(\w_{2j_2})}}-\frac{\overline{F(\b_{1j_1})} F(\w_{2j_2})}{
\overline{\phi(\b_{1j_1})}-\phi(\w_{2j_2})}-
\frac{\overline{F(\b_{1j_1}) F(\w_{2j_2})}}{\overline{\phi(\b_{1j_1})}-
\overline{\phi(\w_{2j_2})}}
\right)
\dots\\
\dots\left(
\frac{F(\b_{kj_k}) F(\w_{1j_1})}{\phi(\b_{1j_1})-\phi(\w_{2j_2})}+
\frac{F(\b_{kj_k})\overline{ F(\w_{1j_1})}}{\phi(\b_{kj_k})-\overline{
\phi(\w_{1j_1})}}-\frac{\overline{F(\b_{kj_k})} F(\w_{1j_1})}{
\overline{\phi(\b_{kj_k})}-
\phi(\w_{1j_1})}-
\frac{\overline{F(\b_{kj_k}) F(\w_{1j_1})}}{\overline{\phi(\b_{kj_k})}-
\overline{\phi(\w_{1j_1})}}
\right)
\end{multline}
}

\noindent plus lower-order terms. 
Multiplying out this product, all $4^k$ terms have an oscillating coefficient
(as some $j_i$ varies)
except for the terms in which the pairs $F(\b_{ij_i})$ and $F(\w_{ij_i})$ 
in the numerator are either both conjugated or both
unconjugated for each $i$. That is, any term involving 
$F(\b_{ij_i})\overline{F(\w_{ij_i})}$
or its conjugate will oscillate as $j_i$ varies and so 
contribute negligibly to the sum.
There are only $2^k$ terms which survive.

Let $z_i$ denote the point $z_i=\phi(\b_{ij_i})\approx\phi(\w_{ij_i})$.

For a term with $F(\b_{ij_i})$ and $F(\w_{ij_i})$ both conjugated,
the coefficient
$$\overline{F(\b_{ij_i}) F(\w_{ij_i})}$$ 
is equal to $d\overline{z_i}$, otherwise it is equal to $dz_i$, where
$dz_i$ is the amount that
$\phi$ changes when moving by one step along path $\gamma_i$, that is,
when $j_i$ increases by $1$.

So the above term for the $k$-cycle $\sigma$ becomes
$$-(\frac{-1}{4\pi i})^k\sum_{\ve_1,\dots,\ve_k=\pm1}\left(\prod_j\ve_j\right)
\int_{\phi(\gamma_1)}\dots\int_{\phi(\gamma_k)}
\frac{dz_1^{(\ve_1)}\dots dz_k^{(\ve_k)}}
{(z_1^{(\ve_1)}-z_2^{(\ve_2)})(z_2^{(\ve_2)}-z_3^{(\ve_3)})\dots
(z_k^{(\ve_k)}-z_1^{(\ve_1)})}$$
plus an error of lower order,
where $z_j^{(1)}=z_j$ and $z_j^{(-1)}=\bar z_j$,
with similar expressions for other $\sigma$.

When we now sum over all permutations $\sigma$, only the fixed-point free
involutions do not cancel:
\begin{lemma}[\cite{Boutillier}]
For $n>2$ let $C_n$ be the set of $n$-cycles in the symmetric group $S_n$. Then
$$\sum_{\sigma\in C_n}\prod_{i=1}^n\frac1{z_{\sigma(i)}-z_{\sigma(i+1)}}=0,$$
where the indices are taken cyclically.
\end{lemma}

\begin{proof}
This is true for $n$ odd by antisymmetry (pair each cycle with
its inverse). For $n$ even,
the left-hand side is a symmetric rational function whose denominator
is the Vandermonde $\prod_{i<j}(z_i-z_j)$ and whose numerator is of lower
degree than the denominator. Since the denominator is antisymmetric,
the numerator must be as well. But the only antisymmetric polynomial of
lower degree than the Vandermonde is $0$.
\end{proof}

By the lemma, in the big determinant all terms cancel
except those for which $\sigma$ is a fixed-point free involution.
It remains to evaluate what happens for a single transposition,
since a general fixed-point free involution 
$\sigma$ will be a disjoint product of these:
{\tiny $$\frac1{(4\pi i)^2}\left[
\int_{\phi(s_1')}^{\phi(s_1)} \int_{\phi(s_2')}^{\phi(s_2)}
\frac{dz_1 dz_2}{(z_1-z_2)^2}-
\int_{\phi(s_1')}^{\phi(s_1)} \int_{\phi(s_2')}^{\phi(s_2)}
\frac{d\bar z_1 dz_2}{(\bar z_1-z_2)^2}-
\int_{\phi(s_1')}^{\phi(s_1)} \int_{\phi(s_2')}^{\phi(s_2)}
\frac{dz_1 d\bar z_2}{(z_1-\bar z_2)^2}+
\int_{\phi(s_1')}^{\phi(s_1)} \int_{\phi(s_2')}^{\phi(s_2)}
\frac{d\bar z_1 d\bar z_2}{(\bar z_1-\bar z_2)^2}
\right]$$}
$$=\frac1{(4\pi i)^2}\int_{\overline{\phi(s_1)}}^{\phi(s_1)}
\int_{\overline{\phi(s_2)}}^{\phi(s_2)}\frac{dz_1 dz_2}{(z_1-z_2)^2}$$
$$=\frac{1}{2\pi}G(\phi(s_1),\phi(s_2))$$
where we used $\phi(s_i')\in\R$.

Now suppose that some of the $s_i$ coincide. We choose paths $\gamma_i$
as before but suppose that the $\gamma_i$ are close only
at those endpoints where the $s_i$ coincide.
Let $\delta=\sqrt{\eps}$.
For pairs of edges $a_{ij_i},a_{i',j_{i'}}$ on different paths, both 
within $\delta$ of such an endpoint we use the bound 
$K^{-1}(\b,\w)=O(\frac1{\eps})$, so that the big
determinant, multiplied by the prefactor $\prod_i K(\w_{ij_i},\b_{ij_i})$,
is $O(1)$. In the sum over paths the net contribution for
each coincidence is then $O(\delta/\eps)^2=O(\eps^{-1})$, since this is the
number of terms in which both edges of two different paths
are near the endpoint.
\end{proof}

\section{Boxed plane partition example}\label{BPPsection}

For the boxed plane partition, whose hexagon has vertices
$$\{(1,0),(1,1),(0,1),(-1,0),(-1,-1),(0,-1)\}$$ in $(\hat x,\hat y)$ coordinates,
see Figure \ref{bpp},
it is shown in \cite{KO} that 
$\Phi$ satisfies
$(-\Phi x+y)^2=1-\Phi+\Phi^2$, or
$$\Phi(x,y)=\frac{1-2xy+\sqrt{4(x^2-xy+y^2)-3}}{2(1-x^2)}$$
for $(x,y)$ inside the circle $x^2-xy+y^2\leq 3/4$. $\Phi$
maps the region inside the inscribed circle with degree $2$ onto the upper half-plane,
with critical point $(0,0)$ mapping to $e^{\pi i/3}$. 
If we map the half plane to the unit disk with the mapping 
$z\mapsto\frac{z-e^{\pi i/3}}{z-e^{-\pi i/3}},$ the composition is
\begin{equation}\label{PhiBPP}
\frac{\Phi(x,y)-e^{\pi i/3}}{\Phi(x,y)-e^{-i\pi/3}}=\frac{2r^2-3+\sqrt{9-12r^2}}{2(y-\bar\omega x)^2}
\end{equation}
(where $r^2=x^2-xy+y^2$)
which 
maps circles concentric about the origin to circles concentric about the origin.
To see this, note that $z=i(y-\omega x)$ defines the standard conformal structure,
and $|z|^2=x^2-xy+y^2=r^2$, so that the right-hand side of the equation
(\ref{PhiBPP}) is $f(r)/\bar z^2$.

Note also that the Beltrami differential of $\Phi$, which is
$$\mu(x,y) = \frac{\Phi_{\bar z}}{\Phi_z} = 
\frac{\Phi-e^{\pi i/3}}{\Phi-e^{-\pi i/3}},$$
satisfies $$\mu=\frac{\mu_{\bar z}}{\mu_z},$$ that is, 
it is its own Beltrami differential!
This is simply a restatement of the PDE (\ref{pde}) in terms of $\mu$.

The diffeomorphism $\phi$ from the region inside the 
inscribed circle to the unit disk is also very simple,
it is just $\phi=\sqrt{\mu},$ or
$$\phi(re^{i\theta})=\frac{\sqrt{3}-\sqrt{3-4r^2}}{2r}e^{i\theta}.$$
The inverse of $\phi$ which maps $\D$ to $U$ is even simpler: it is
$$\phi^{-1}(z) = \frac{z\sqrt{3}}{1+|z|^2}.$$
This map can be viewed as the orthogonal projection of a hemisphere onto
the plane through its equator, if we identify $\D$ conformally with the upper
hemisphere sending $0$ to the north pole.

In conclusion if the domain $U$ is the disk
$\{(x,y)~|~x^2-xy+y^2\leq r^2\}$, where $r^2<3/4$, 
and the height function on the boundary of $U$
is given by the height function $\bh$ of the BPP on $\partial U$, then the
$\bh$ on $U$ will equal the $\bh$ on BPP restricted to $U$, and
the fluctuations of the height function are the pull-back of the Gaussian
free field on the disk of radius $\frac{\sqrt{3}-\sqrt{3-4r^2}}{2r}$
under $\phi$.

\end{document}